\documentclass[a4paper,11pt]{article}
\usepackage{jcappub}
\usepackage[T1]{fontenc}
\usepackage{ae,aecompl}
\usepackage{graphicx}
\usepackage{amsmath}
\usepackage{amssymb}
\usepackage{lipsum}
\usepackage{mathtools}
\usepackage{comment}
\usepackage{multirow}
\usepackage{xcolor}
\usepackage{subfigure}
\usepackage{hyperref}
\usepackage{latexsym}
\usepackage{pifont}
\usepackage{adjustbox}
\usepackage{soul}
\usepackage[normalem]{ulem}
\graphicspath{{./Pictures/}}
\pdfoutput=1

\def\mean#1{\left< #1 \right>}          
\newcommand{\tvec}[1]{\mathbf{#1}}      
\newcommand{\de}{\,\mathrm{d}}          
\newcommand{\sub}{\mathrm}

\newcommand{\derp}[2]{\frac{\partial #1}{\partial #2}}

\newcommand{\xmark}{\ding{55}}

\newcommand{\di}{\text{d}}

\title{\boldmath  The effects of massive neutrinos on the linear point of the correlation function}

\author[a,b,c]{G.\ Parimbelli}
\emailAdd{gabriele.parimbelli@sissa.it}

\author[d,e]{, S.\ Anselmi}
\author[a,c,f,b]{, M.\ Viel}
\author[g]{, C.\ Carbone}
\author[h]{, F.\ Villaescusa-Navarro}
\author[e,i]{, P.S.\ Corasaniti}
\author[e]{, Y. Rasera}
\author[j,k]{, R. Sheth}
\author[l]{, G.D.\ Starkman}
\author[l]{and I. Zehavi}

\affiliation[a]{SISSA - International School for Advanced Studies, Via Bonomea 265, 34136 Trieste, Italy}
\affiliation[b]{INFN - National Institute for Nuclear Physics, Via Valerio 2, 34127 Trieste, Italy}
\affiliation[c]{IFPU - Institute for Fundamental Physics of the Universe, Via Beirut 2, 34151 Trieste, Italy}
\affiliation[d]{TECHNION - Department of Physics - Israel Institute of Technology, Haifa 320003 -- Israel}
\affiliation[e]{LUTH - UMR 8102 CNRS, Observatoire de Paris, PSL Research University, Universit\'e de Paris, 92190 Meudon -- France}
\affiliation[f]{INAF-OATS, via Tiepolo 11, 34131 Trieste, Italy}
\affiliation[g]{INAF-IASF Milano, Via Alfonso Corti 12, I-20133 Milano, Italy}
\affiliation[h]{Department of Astrophysical Sciences, Princeton University, Princeton, New Jersey 08544, USA}
\affiliation[i]{Institut d'Astrophysique de Paris, CNRS UMR 7095 and UPMC, 98bis, bd Arago, F-75014 Paris -- France}
\affiliation[j]{Center for Particle Cosmology, University of Pennsylvania, 209 S. 33rd St., Philadelphia, PA 19104 -- USA}
\affiliation[k]{The Abdus Salam International Center for Theoretical Physics, Strada Costiera, 11, Trieste 34151 -- Italy}
\affiliation[l]{Department of Physics/CERCA/Institute for the Science of Origins, Case Western Reserve University, Cleveland, OH 44106-7079 -- USA}

\abstract{
The linear point (LP), defined as the mid-point between the dip and the peak of the two-point clustering correlation function (TPCF), has been shown to be an excellent standard ruler for cosmology.
In fact, it is nearly redshift-independent, being weakly sensitive to non-linearities, scale-dependent halo bias and redshift-space distortions. So far, these findings were tested assuming that neutrinos are massless; in this paper we extend the analysis to massive-neutrino cosmologies.
In particular, we examine if the scale-dependent growth induced by neutrinos affects the LP position and if it is possible to detect the neutrino masses using the shift of the LP compared to the massless-neutrino case.
For our purposes, we employ two sets of state-of-the-art $N$-body simulations with massive neutrinos. For each of them we measure the TPCF of cold dark matter (CDM) and halos and, to estimate the LP, fit the TPCF with a model-independent parametric fit in the range of scales of the Baryon Acoustic Oscillations (BAO).
Overall, we find that the LP retains its features as a standard ruler even when neutrinos are massive. The cosmic distances measured with the LP can therefore be employed to constrain the neutrino mass.
}

\begin{document}

\maketitle
\flushbottom


\section{Introduction}
\label{sec:introduction}
In the early stages of the Universe, photons and baryons were tightly coupled in a single primordial plasma kept at thermal equilibrium by Thomson scattering.
The opposing effects of gravity and radiation pressure produced propagating acoustic waves.
When the Universe became too cold and dilute, this interaction ceased to effectively couple the photons and baryons, and the acoustic waves ceased to propagate.
Nevertheless, these oscillations left an imprint on the distribution of matter, with a characteristic scale of approximately the sound horizon, \emph{i.e.} the distance the waves had propagated.

Today we can observe the evidence of this interaction, the so-called Baryon Acoustic Oscillations (BAO), either in Fourier space, as wiggles in the power spectrum of matter or its tracers, or in configuration space as a peak in the 2-point clustering correlation function (TPCF)  \cite{2dF_BAO-Cole+05, SDSS_BAO-Eisenstein+05}.
More recently, full-shape analyses of the TPCF have been performed for cosmological parameter estimation \cite{2012MNRAS.425..415S}, while several works with the same aim have been carried out in Fourier space \cite{2009MNRAS.400.1643S,Sanchez-BOSS+13,2017MNRAS.464.1640S, ivanov,philcox} also in relation to neutrino physics \cite{6,7,8,9,10}.

BAOs became an important tool in cosmology because in principle they provide a powerful standard ruler: they have been shown to be very robust against systematics (see e.g. Ref. \cite{BAO_systematics_BOSS-Ross+17}) that critically affect other observables, like the full-shape power spectrum.
They allow us to measure the acoustic scale, a quantity that is independent of the spatial geometry of the Universe, the primordial fluctuation parameters, late-time acceleration and the choice of observed tracers (e.g. galaxies) of the underlying density field.
In other words, BAOs can be used to map the expansion history of the Universe through estimates of the Hubble parameter and the angular-diameter distance, exploiting the Alcock-Paczynski (AP) distortions \cite{AP79} (usually called cosmic distance estimates).
Unfortunately, there are a few effects that make the usage of BAOs difficult:
non-linearities in the late Universe affect the TPCF and in particular the position of the BAO peak \cite{BAO-Desjaques+10, BAO_evolution-Baldauf+17, BBKS+86} (the originally proposed BAO standard ruler \cite{SDSS_BAO-Eisenstein+05}). This  spoils the standard ruler nature of the peak \cite{BAO_peak_motion-Smith+08, Best_way_to_measure_BAO-Sanchez+08}.

Current analyses circumvent this problem by fitting the data with a theoretical template of the TPCF parametrized in terms of the linear TPCF.
The most widely used method to estimate cosmic distances from the TPCF in the BAO region (see e.g. Refs. \cite{BAO_nonlinear-Seo+08, 2-per-cent-SDSS-II+12, BOSS_DR10-11-Anderson+14}) consists of fixing the cosmological parameters to the fiducial vanilla-$\Lambda$CDM values used to generate the mock catalogs from which the covariance matrix is computed. The non-linear damping parameter is also estimated from the mocks and kept fixed in the MCMC analysis, checking \textit{a posteriori} that it does not affect the measurement of the AP distortion parameter.
This method has been shown to accurately fit the TPCF 
and to return unbiased distance measures.
Unfortunately, it might suffer from some drawbacks.
First, the value of the damping parameter is tracer dependent \cite{BBKS+86, Distance-z_relation-Veropalumbo+16}, and fixing it leads to unjustified claims for precision and accuracy.
These assumptions could result in an underestimate of the distance error.
Second, mock catalogs are typically generated using a  vanilla-$\Lambda$CDM model, therefore it is also not precisely clear how these measurements apply to non-standard cosmology scenarios (e.g. non-flat geometries and evolving dark energy).
All in all, employing this method might underestimate the distance errors by up to a factor of 2 \cite{Linear_point_IV-BAO-Anselmi+18}.

In recent years, a new potential standard ruler, dubbed the linear point (LP), has been proposed \cite{Linear_point_I-theory-Anselmi+15,Linear_point_II-BOSS-Anselmi+17,Linear_point_III-method-Anselmi+17,Linear_point_IV-BAO-Anselmi+18,2020PhRvD.101h3517O}: the LP is defined as the mid-point between the BAO peak and the dip of the TPCF.
The LP has been shown to be weakly affected by  gravitational processes.
First of all, it is insensitive to the primordial fluctuation amplitude $A_\mathrm s$ and the scalar spectral index $n_\mathrm s$ \cite{Linear_point_I-theory-Anselmi+15,2020PhRvD.101h3517O}.
Second, the original analysis of Ref.~\cite{Linear_point_I-theory-Anselmi+15} found that late-time non-linearities move the BAO peak towards smaller scales and the dip in the opposite direction, thus leaving the LP nearly in the same position.
Similarly, redshift-space distortions (RSD) do not influence the position of the LP, as their effect on the peak and the dip nearly cancels.
Finally, the position of the LP is also nearly unaffected by scale-dependent halo bias.
The stability of the LP appears to be due to the near-antisymmetry of the TPCF with respect to the LP itself, limiting the downward drift of the LP between high and low redshifts to $\sim1\%$. Given the secular nature of that shift, to partially remove this non-linear effect, Ref. \cite{Linear_point_I-theory-Anselmi+15} introduced a simple redshift-independent 0.5\% correction to the LP estimated from real or simulated data:

\begin{equation}
    r_\sub{LP} = \frac{r_\sub{d}+r_\sub{p}}{2}\times 1.005.
    \label{eq:LP_definition}
\end{equation}

The LP is identified by first finding the dip and the peak and it is subsequently used to estimate the isotropic volume distance to the redshift considered.
Finally, the value of such distance is compared to the theoretical predictions from different models in order to constrain the cosmological parameters.

Given what said above, a cosmological-model-independent fit is sufficient to recover the LP position without introducing systematics.
LP analyses thus employ a model-independent approach to estimate the LP from real or simulated clustering data. In particular, it has been found that a simple polynomial fit is enough to obtain an unbiased estimate for the LP \cite{Linear_point_III-method-Anselmi+17}, with a correct propagation of the uncertainties.

So far, the LP has been tested only in the vanilla-$\Lambda$CDM framework, with no investigation of the possible impact of massive neutrinos.
Massive neutrinos affect the clustering of matter both at the linear and non-linear levels.
They decoupled from the baryon-photon plasma in the very early Universe, when they were still relativistic particles.
Due to their high thermal velocities, they cannot cluster, at linear order, on regions smaller than their so-called free-streaming scale (see e.g. Ref. \cite{Lesgourgues-Pastor+12}):
\begin{equation}
\lambda_\sub{fs} = 8 \ (1+z) \ \frac{H_0}{H(z)} \frac{1 \ \textrm{eV}}{m_\nu} \ h^{-1} \ \textrm{Mpc},
\label{eq:free-streaming}
\end{equation}

where $m_\nu$ is the mass of a single neutrino species (as opposed to $M_\nu\equiv\sum m_\nu$, which will label the sum of neutrino masses).

Notice that, for particles becoming non-relativistic during matter domination, as is usually the case for massive neutrinos, the {\em comoving} free-streaming length, $\lambda_\sub{fs}/a$, decreases in time, and therefore achieves its maximum value at the time of the non-relativistic transition. This free-streaming distance corresponds to the wave-number
\begin{equation}
k_\sub{nr} = k_\sub{fs} (z_\sub{nr})\simeq 0.018\,\Omega_m^{1/2}\,\left(\frac{1\,{\rm eV}}{m_\nu}\right)^{1/2}\ h/\textrm{Mpc}\,,
\end{equation}
\emph{i.e.} to a scale typically larger than the scale at which nonlinear effects manifest themselves at low redshifts.
At any redshift, massive neutrinos prevent the growth of structures on $k>k_\sub{fs}(m_{\nu},z)\equiv 2\pi a/\lambda_{\rm fs}(m_\nu,z)$; on the other hand, scales larger than $1/k_\sub{nr}$ are never affected by free-streaming and in that regime neutrinos behave exactly like CDM.
As a result, the linear growth of density fluctuations becomes scale-dependent, with a substantial impact on the matter power spectrum and the TPCF.
Therefore, the position of the peak, dip and LP in the TPCF could be affected by the value of the neutrino masses, even before the effect of gravitational non-linearities. 

In this paper we quantify the impact of massive neutrinos on the LP. We start by studying how the scale-dependence clustering induced by massive neutrinos impacts the peak, dip and LP positions in linear theory. We then investigate the effects of non-linearities by using state-of-the-art $N$-body simulations.
We focus on the behavior of the LP in the TPCF for both cold dark matter (CDM) and halos in real space, leaving the analysis of redshift-space for future work.
The goal of this work is to investigate whether the neutrino mass retains or spoils the features of the LP that are crucial when employing it as a standard ruler. Given our encouraging findings, we discuss how the LP could be applied to constrain the cosmological energy densities and the neutrino masses.

The paper is organized as follows.
In Section \ref{sec:method} we develop the methodology we employ, \emph{i.e.} the simulation sets and the LP estimation procedure. In Section \ref{sec:results} we present and discuss our results. Finally, in Section \ref{sec:conclusions} we draw our conclusions.


\section{Methodology}
\label{sec:method}

The goal of this paper is to study the evolution of the LP through cosmic ages, in particular addressing the influence of massive neutrinos on the LP position. To this end, we first investigate the effect of the scale-dependent growth in linear theory.
We then move to the non-linear analysis, employing $N$-body simulations that incorporate massive neutrinos as an extra set of particles.
We use as observables the CDM and halo TPCF in real space, leaving the impact of RSD for future work. In order to estimate the LP position, we fit each of these TPCFs with a cosmology-independent polynomial function. Note that we do not consider the CDM-plus-neutrinos TPCF as an observable; because of the large neutrino free-streaming scale, it was shown \cite{Ichiki+11,nuLCDM1+13,nuLCDM2+13} that the main driver of galaxy formation is the cold dark matter plus baryons component rather than total matter. 
Using the total matter density as the fundamental field would spoil the universality of the mass function (proven by Refs. \cite{universal_hmf-Jenkins+01, universal_hmf-Reed+03}) and would give rise to a scale-dependent halo and galaxy bias at the largest scales.
Therefore the CDM-plus-baryons field is expected to be the closest underlying field of the tracers we observe in the Universe\footnote{From here on, when mentioning the CDM field, we will refer to the CDM-plus-baryons one.}.
Furthermore, Ref. \cite{Vagnozzi+18} showed that, in upcoming surveys, not accounting for a scale-dependent bias when using total matter as fundamental field will lead to substantial shifts in the posterior of $M_\nu$ as well as of other cosmological parameters which are correlated with it.

In this Section, we first explain how we perform the linear analysis, then present the adopted simulation set, together with the method used to measure the correlation functions for both CDM and halos.
We next describe the procedure we followed to estimate the LP best fit and uncertainty, fitting a model-independent differentiable function to the TPCF data and errors. We also compare the TPCF $N$-body-estimated covariance matrix to its linear Gaussian prediction.


\subsection{Effects of massive neutrinos in linear theory}
\label{sec:linear}

The fundamental feature that makes the LP a standard ruler is that its position is nearly redshift-independent in comoving coordinates \cite{Linear_point_I-theory-Anselmi+15,Linear_point_IV-BAO-Anselmi+18}. Ref.~\cite{Linear_point_IV-BAO-Anselmi+18} explained that the LP can be used to estimated cosmological distances for vanilla-$\Lambda$CDM, and for cosmological models that do not introduce a scale-dependent growth, i.e.~cosmologies that retain the LP redshift-independence. In order to understand whether the LP is a standard ruler for massive-neutrino cosmologies we must assess the impact of massive neutrinos on the LP position, first in linear theory and then taking into account late-time non-linearities.

{To investigate the redshift-dependence of the LP in linear theory for different neutrino masses, we use a Boltzmann code (e.g.~the CLASS code \cite{2011JCAP...09..032L}). We obtain the linear CDM
power spectrum $P_\mathrm c^{\rm lin}(k, z)$ at redshift $z$ from CLASS, and compute the spatial derivative of the real-space TPCF through:
\begin{equation} \label{xi:deriv}
	\xi^{{\prime}}(r, z) = -\frac{1}{2\pi^2} \int \di k \,\,k^{3}P_\mathrm c^{\rm lin}(k, z) j_{1}(k r) \,,
\end{equation}
where $j_{1}(x)= (- x \cos(x) + \sin(x))/x^{2}$ is the first-order spherical Bessel function. 
We calculate the dip, peak and linear point positions by applying a root-finding routine to the condition  $\xi^{{\prime}}(r, z) = 0$ (without the 0.5\% correction mentioned in Eq.~(\ref{eq:LP_definition})).


\subsection{Simulations}
\label{sec:sims}

In this work, we employ two sets of $N$-body simulations with massive neutrinos. As usual in $N$-body simulations, baryons are treated as cold dark matter, hence the CDM $N$-body particles are meant to describe the cold dark matter plus baryons component.

The first simulation suite is a new subset of the ``Dark Energy and Massive Neutrino Universe'' (DEMNUni) simulations, first presented in Refs. \cite{DEMNUni+16,DEMNUni-Carbone+16}.
The complete DEMNUni set encloses simulations with different cosmologies and can be regarded as the state-of-the-art simulations in terms of both mass resolution and volume \cite{Schuster, Verza,  Kreisch, Bel, Ruggeri+18}.
This new suite consists of 50 realizations of two different models, vanilla-$\Lambda$CDM, and a $\nu\Lambda$CDM with three degenerate neutrino species of total mass $M_\nu = 0.16$ eV.
The other parameters are set to $\Omega_\sub m = 0.32$, $\Omega_\sub b = 0.05$ $h = 0.67$, $n_\sub s = 0.96$, $A_\sub s = 2.1265 \times 10^{-9}$.
The latter parameter implies a value for $\sigma_8 = 0.833$ and $0.792 $ for the $\Lambda$CDM and for the massive neutrino cases, respectively.

The new DEMNUni set, considered in this work, has been run using the tree-particle mesh-smoothed particle hydrodynamics (TreePM-SPH) code Gadget-III \cite{Springel+05}, opportunely modified  \cite{viel+10} to account for the presence of massive neutrinos.
The simulation follows the gravitational evolution -- starting from an initial redshift of $z_\sub{in} = 99$ to the present age -- of $N_c = 1024^3$ CDM particles and, when present, $N_\nu=1024^3$ neutrino particles, in a cubic box of size $L=1000$ Mpc/$h$.
Initial conditions for models with massive neutrinos are obtained via the rescaling method developed in Ref. \cite{Zennaro+17}.
With the cosmological parameters above, the mass of a CDM particle is $M_P^\sub c \approx 8.2\times 10^{10} \ M_\odot/h$, while when neutrinos are present each particle has a mass of $M_P^\nu \approx 9.9\times 10^{8} M_\odot/h$.
The softening length has been set to $\varepsilon=20 \ h^{-1}$ kpc.
With these features, DEMNUni are suitable for the analysis of several cosmological probes, from galaxy clustering to weak lensing.
Halos and sub-halos are identified via the Friends-of-Friends (FoF) and the SUBFIND algorithms respectively, both included in Gadget-III \cite{GADGET+01, SUBFIND+08}, setting the linking length to $1/5$ of the mean inter-particle separation.
The minimum number of particles to identify a parent halo is $32$, so that the minimum halo mass is $2.6\times 10^{12} \ M_\odot/h$.
In this work we consider 5 snapshots at $z=0, 0.5, 1, 1.5, 2$.

We also employ part of the new Quijote set \cite{Quijote+19}.
Like the DEMNUni, these simulations are run with the TreePM code Gadget-III.
However, the initial conditions are set at $z_\sub{in} = 127$ (also here using the presciption by Ref. \cite{Zennaro+17}), and the mass resolution is 8 times lower, with $N_\sub c=512^3$ CDM particles, and $N_\nu=512^3$ neutrinos (when present), in a box of 1000 Mpc/$h$ on each side.
The fiducial cosmology of this set has $\Omega_\sub m = 0.3175$, $\Omega_\sub b = 0.049$, $h = 0.6711$, $n_\sub s = 0.9624$, $\sigma_8 = 0.834$.
Neutrinos are considered to be of three different species  with degenerate masses.
This means that a dark matter particle has a mass of $M_P^\sub{c} \approx 6.5\times 10^{11} M_\odot/h$, while neutrino particles have $M_P^\sub{\nu} \approx 1.6\times 10^{10} M_\odot/h \ \times M_\nu \ [\mathrm{eV}]$.
Dark matter halos, with a minimum mass of $1.3\times 10^{13} M_\odot/h$ (32 CDM particles), are identified through the FoF algorithm with linking length parameter set to $1/5$ of the mean inter-particle separation.
Also in this case, we use 5 different snapshots at $z=0, 0.5, 1, 2, 3$.

Given the number of Quijote realizations available, we do not use the full set.
We included just the first 500 realization of the vanilla-$\Lambda$CDM model plus the 500 standard realizations corresponding to a neutrino mass sum of $0.1$ and $0.2$ eV. We do not use at all the 500 realizations with $M_\nu = 0.4$ eV.
This is because in the Quijote set, the amplitude of the power spectrum is described by $\sigma_8$ (rather than $A_\sub s$), which is kept fixed to $0.834$.
When $M_\nu = 0.4$ eV, the large-scale amplitude is so large that late-time non-linearities completely smear out the BAO peak in the TPCF, making our analysis impossible to perform.

These 500 simulations have been used to test the accuracy of an analytic Gaussian covariance matrix (see e.g. Ref. \cite{Covmat-Gaussian-Grieb+15}), while the LP estimation procedure has been performed only on the first 100 realizations.

Tables \ref{tab:sims} and \ref{tab:halos_in_sims} summarize the specifics of the two simulations sets just described, and the average number of halos per realization (or equivalently per (Gpc/$h$)$^3$).

\begin{table}
\begin{center}
\renewcommand{\arraystretch}{1.25}
\begin{tabular}{ l||c||c } 
                                                       & \textbf{DEMNUni}     & \textbf{Quijote}    \\
 \hline
 \textbf{Realizations (per model)}                     & 50                   & 100                  \\ 
 \textbf{Boxsize (Mpc/$\boldsymbol h$)}                & 1000                 & 1000                \\ 
 \textbf{Snapshots ($\boldsymbol z$)}                  & 0, 0.5, 1, 1.5, 2    & 0, 0.5, 1, 2, 3     \\ 
 \textbf{CDM particles}                                & $1024^3$             & $512^3$             \\ 
 \textbf{Neutrino particles}                           & $1024^3$             & $512^3$             \\ 
 \textbf{Neutrino mass (eV)}                           & 0, 0.16              & 0, 0.1, 0.2         \\
 \textbf{Minimum halo mass $\boldsymbol{(M_\odot/h)}$} & $2.6\times 10^{12}$  & $1.3\times 10^{13}$ \\ 
\end{tabular}
\end{center}
\caption{Different specifics of the two simulation sets employed in this paper.}
\label{tab:sims}
\end{table}

\begin{table}
\begin{center}
\renewcommand{\arraystretch}{1.25}
\begin{tabular}{ l||c|c||c|c|c } 
 \multirow{2}{*}{} & \multicolumn{2}{c||}{\textbf{\# Halos: DEMNUni}} & \multicolumn{3}{c}{\textbf{\# Halos: Quijote}} \\
  \cline{2-6}
    & $\boldsymbol{\Lambda}$\textbf{CDM} & \textbf{0.16 eV} & $\boldsymbol{\Lambda}$\textbf{CDM} & \textbf{0.1 eV} &  \textbf{0.2 eV} \\
 \hline
 \textbf{$z=3$}   
    &                 &                 & $4.9\times10^3$ & $5.0\times10^3$ & $5.5\times10^3$ \\
 \textbf{$z=2$}   
    & $6.1\times10^5$ & $5.5\times10^5$ & $4.4\times10^4$ & $4.4\times10^4$ & $4.4\times10^4$ \\ 
 \textbf{$z=1.5$} 
    & $1.1\times10^6$ & $9.9\times10^5$ &                 &                 &                 \\
 \textbf{$z=1$}   
    & $1.4\times10^6$ & $1.3\times10^6$ & $2.0\times10^5$ & $2.0\times10^5$ & $2.0\times10^5$ \\
 \textbf{$z=0.5$} 
    & $1.8\times10^6$ & $1.7\times10^6$ & $3.1\times10^5$ & $3.1\times10^5$ & $3.1\times10^5$ \\
 \textbf{$z=0$}   
    & $1.9\times10^6$ & $1.9\times10^6$ & $4.1\times10^5$ & $4.1\times10^5$ & $4.1\times10^5$ \\
 \end{tabular}
\end{center}
\caption{Average number of halos per realization, per snapshot and simulation set.}
\label{tab:halos_in_sims}
\end{table}


\subsection{Estimating the TPCF from simulations}
\label{sec:tpcf}

For each snapshot and each realization, we compute the TPCF for CDM-only and for halos.
As explained at the beginning of Section \ref{sec:method}, we exclude neutrino particles from the computation of the TPCF for observational reasons. This is also convenient from the theoretical point of view -- as widely explained in e.g. Refs. \cite{nuLCDM1+13, nuLCDM2+13, nuLCDM3+13}, in massive-neutrino cosmologies, if we consider the CDM density field we obtain a universal halo mass function and an \emph{almost} scale-independent linear halo bias.
We recall that we limit ourselves to the real space TPCF, leaving RSD analysis for future work.

The TPCF is computed using the FFT estimator introduced in Ref. \cite{Taruya+09} and implemented in  the \textsc{Pylians} codes \footnote{\url{https://github.com/franciscovillaescusa/Pylians}}, in which the density field is computed on a grid and convolved with itself through a double 3-dimensional Fast Fourier Transform (FFT):

\begin{equation}
\hat{\xi}_X^\sub{sim}(r) = \frac{1}{N_\sub{modes}} \sum_{r_\sub{min}<|\tvec r| < r_\sub{max}} \mathrm{FFT}^{-1} \left[ \left|\delta_X(\tvec k)\right|^2\right](\tvec r),
\label{eq:Taruya_estimator}
\end{equation}

where $X$ can be either `c' for CDM or `h' for halos.
The density field $\delta_X$ is computed using a Cloud-In-Cell mass-assignment scheme. 
The bin edges $r_\sub{min}$ and $r_\sub{max}$ are fixed by the thickness of the grid: in this work we set the latter to 1024, corresponding to a bin size of roughly 1 Mpc/$h$.

The FFT estimator allows one to considerably reduce the computational time compared to brute-force particle pair counts (unfeasible given the number of particles in our simulations). The available convergence studies made on the FFT estimator do not yet allow one to assess whether, in the BAO range-of-scales, the FFT-estimated TPCF reaches a $1\%$ level of accuracy \cite{Taruya+09}. While  detailed investigation of the FFT estimator numerical issues will be the subject of future work, in Appendix \ref{app:numerical_systematics} we explain the methodology we developed to minimize its impact on LP estimation.

In Fig.~\ref{fig:DEMNUni_Quijote_TPCF} we plot the TPCF of CDM as measured with the method just described.
For an easier comparison, we plot the quantity $r^2\xi(r)/\sigma_8^2 D_1^2(z)$ for the redshifts in common between the two simulation sets.
We split the measurements for vanilla-$\Lambda$CDM from the ones with massive neutrinos.
Dark-red dots and dark-blue dots in the left-hand panels represent the TPCF for the $\Lambda$CDM model of the DEMNUni and the Quijote simulations, respectively.
In the right-hand panels we show the massive-neutrino models of the DEMNUni in light red and of the Quijote in blue (for $M_\nu = 0.1$ eV) and light blue ($0.2$ eV).
Each measurement is accompanied by the standard error on the mean as uncertainty.

We would like to underline a subtle difference between the measurements of the two sets.
As already written in Section \ref{sec:sims}, in the Quijote simulations the parameter ruling the overall amplitude is $\sigma_8$ and not $A_\mathrm s$.
Therefore we expect a larger flattening of the BAO feature at late times for a fixed neutrino mass.
This is clearly visible in Fig.~\ref{fig:DEMNUni_Quijote_TPCF}, where (for instance in the Quijote TPCF for $M_\nu = 0.1$ eV -- blue points in the right panels) the relative height between the dip and the peak is smaller than for the analogous DEMNUni TPCF, despite in the latter the neutrino mass is even higher (0.16 eV).
This has important consequences on the estimate of the LP and in particular of its uncertainty.
We will discuss this in detail in Section \ref{sec:error-forecast}.

\begin{figure}[t]
    \centering
    \includegraphics[width=1.\textwidth]{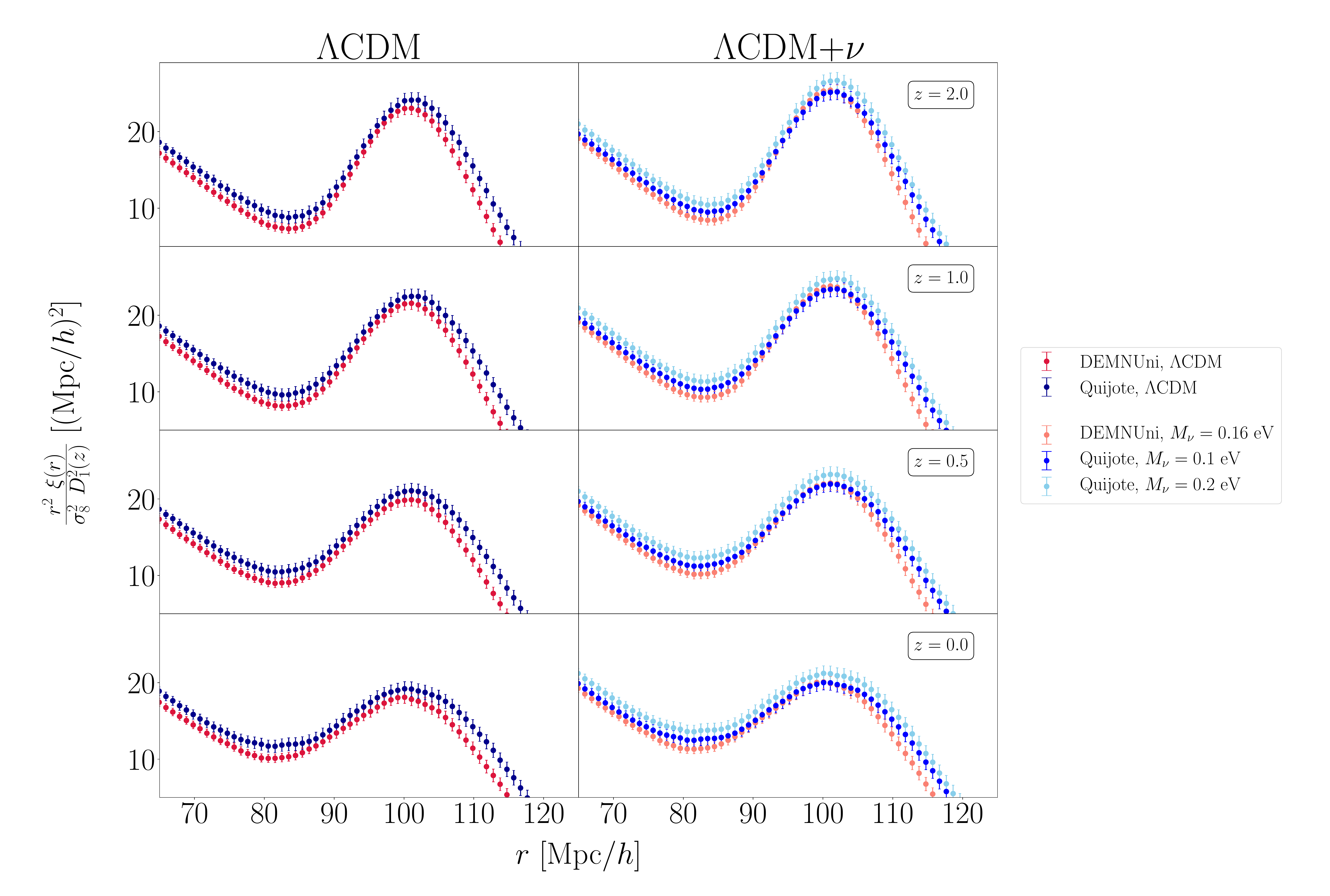}
    \caption{TPCF of CDM from the DEMNUni and the Quijote sets, as measured with Eq.~(\ref{eq:Taruya_estimator}). We show here, only for the common redshift between the two sets, the TPCF multiplied by $r^2$ and divided by the $\sigma_8^2$ and the growth factor for $\Lambda$CDM squared in order to make it easier a comparison between different sets.
    To facilitate the comparison, we plot the mean of the 50 DEMNUni and of the first 50 Quijote realizations, each with an uncertainty corresponding to the standard error on the mean.
    In the left panels we display the two vanilla-$\Lambda$CDM cases, with the DEMNUni in dark red and the Quijote in dark blue; the right panels are left for the massive neutrino models, with the DEMNUni in light red and the Quijote with 0.1 eV (0.2 eV) in blue (light blue).}
    \label{fig:DEMNUni_Quijote_TPCF}
\end{figure}

\begin{figure}
\includegraphics[width=\textwidth]{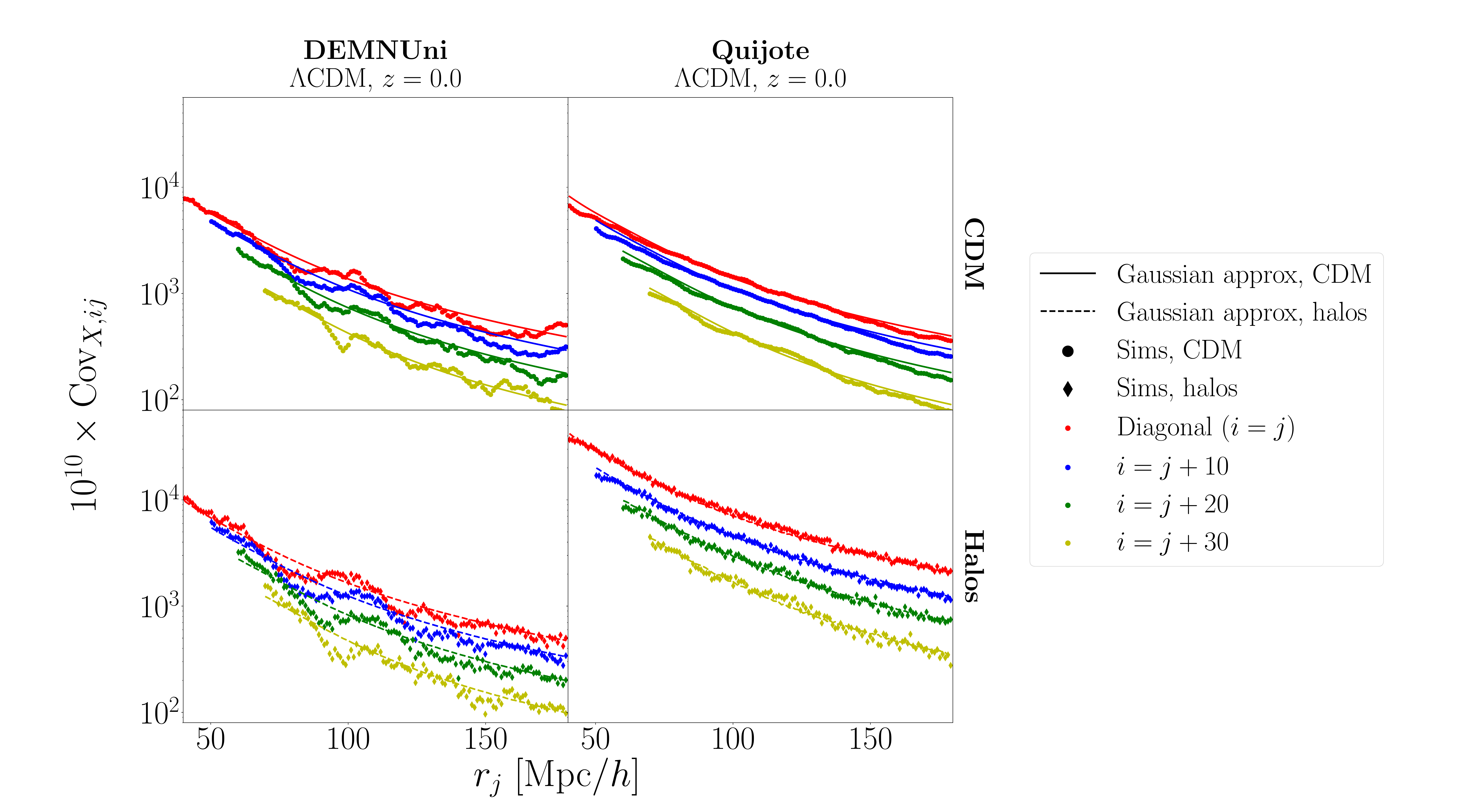}
\caption{Covariance of the TPCF in our simulations for the vanilla-$\Lambda$CDM at $z=0$.
The left panels refer to the DEMNUni set, while the right panels show the same but for the Quijote simulations. Here we rescale the covariance by the number of realizations, \emph{i.e.} we represent the covariance of the TPCF in a cubic box of side 1000 Mpc/$h$. Dots and diamonds represent the measured covariance of CDM (top panels) and halos (bottom), respectively, while solid and dashed lines are the analytical equivalent under the assumption of a Gaussian density field (see Eq.~(\ref{eq:covmat})). Different colors label different elements of the covariance matrix: red is for the diagonal elements (\emph{i.e.} the variance of the TPCF), while blue, green and yellow show respectively the 10-th, 20-th and 30-th off-diagonal elements (with an offset introduced for sake of clarity).}
\label{fig:covariance_lcdm}
\end{figure}


\subsection{Estimating the linear point from simulations}
\label{sec:lp}

In Ref.~\cite{Linear_point_III-method-Anselmi+17} it was shown that the LP position can be extracted from $N$-body simulations, mock and real galaxy data in a cosmology model-independent way. The proposed procedure exploits a simple polynomial function to smooth the binned TPCF data and estimate the zero-crossings of its first derivative. This polynomial function is written as:
\begin{equation}
   \xi^\sub{fit}_ X(r) = \sum_{n=0}^N a_n r^n 
    \,
   \label{eq:model_TPCF}
\end{equation}
where the degree of the polynomial $N$ must be chosen following Ref. \cite{Linear_point_III-method-Anselmi+17}.
The best-fit parameters are found by maximizing the log-likelihood function given by:
\begin{equation}
    \ln \mathcal{L} \propto - \frac{1}{2} \sum_{i,j} \left[\xi^\sub{fit}_X (r_i)-\hat{\xi}^\sub{sim}_X (r_i)\right] \left[\textrm{Cov}^{-1}_X\right]_{ij} \left[\xi^\sub{fit}_X (r_j)-\hat{\xi}^\sub{sim}_X (r_j)\right],
    \label{eq:likelihood}
\end{equation}

where $\hat{\xi}^\sub{sim}_X(r)$ is the correlation function estimated from the simulation for either CDM or halos ($X={\rm c, h}$), Eq.~(\ref{eq:Taruya_estimator}), $\textrm{Cov}_{X,ij}$ is the corresponding covariance matrix, and $\xi^\sub{fit}_X(r)$ is the polynomial employed to estimate the LP.

We recall that, following Eq.~(\ref{eq:LP_definition}), the LP is defined as the mid-point between the peak and the dip in the TPCF
(plus a $0.5$\% correction).
We need to propagate the uncertainty from the fitted parameters of the TPCF to the position of the peak and the dip, and finally to the  LP.
To do so, we write the LP position as a function of the polynomial coefficients of Eq.~(\ref{eq:model_TPCF}), and expand the result in the vicinity of the best-fit parameters.
Assuming that the uncertainties in the $a_i$'s are small (as we verify numerically), we can stop at first order:

\begin{equation}
    r_\sub{LP}(\boldsymbol a) \approx  r_\sub{LP}(\boldsymbol{\bar a}) + \sum_i \derp{r_\sub{LP}(\boldsymbol{\bar a})}{a_i} \ (a_i - \bar a_{i}).
    \label{eq:LP_taylor}
\end{equation}

The error on the LP is the variance of the first-order term, namely

\begin{equation}
    \sigma_\sub{LP} = \left\{ \sum_{i,j} \frac{\partial r_\sub{LP}}{\partial a_i} \left[\textrm{Cov}(\boldsymbol{\bar a})\right]_{ij} \frac{\partial r_\sub{LP}}{\partial a_j}\right\}^{1/2},
    \label{eq:LP_error}
\end{equation}

where $\textrm{Cov}(\boldsymbol{\bar a}) = \mean{(a_i - \mean{a_i})(a_j - \mean{a_j})}$ is the covariance matrix of the parameters.
The derivative of the LP position with respect to the parameters is computed with the 5-point stencil method:
\begin{equation}
\frac{\partial r_\sub{LP}}{\partial a_i}
\approx 
\frac{-r_\sub{LP}(a_i+2\epsilon) + 8 r_\sub{LP}(a_i+\epsilon) - 8 r_\sub{LP}(a_i-\epsilon) + r_\sub{LP}(a_i-2\epsilon)}{12\epsilon} .
\label{eq:LP_derivative}
\end{equation}

The step $\epsilon$ must be taken in such a way as to guarantee numerical convergence of the derivative.
For every case, we choose $\epsilon = \{10^{-6}, 10^{-7}, 10^{-8}\} \ a_i$ as step sizes, and compare the resulting errors on the LP.
If, for each choice of $\epsilon$, the computed numerical derivatives agree within 1\%, we say that the derivative has converged and we take the value for $\sigma_\sub{LP}$ corresponding to $\epsilon=10^{-7} a_i$. 
When performing this operation, $\textrm{Cov}(\boldsymbol{\bar a})$ is kept fixed, since the mean values of the $a_i$ do not change in this procedure.

The TPCF covariance matrix, needed to minimize the log-likelihood defined by Eq.~(\ref{eq:likelihood}), is computed analytically, as the one computed directly from either set of simulations would be noisy.
Since the scales we are interested in are just mildly non-linear, we can still use a Gaussian-density-field approximation to build the TPCF covariance matrix.
We follow the treatment developed, employed, and tested in Refs. \cite{BAO_peak_motion-Smith+08,Best_way_to_measure_BAO-Sanchez+08,2009MNRAS.400..851S,2-per-cent-SDSS-II+12,Covmat-Gaussian-Grieb+15,2019MNRAS.482.1786L,Linear_point_IV-BAO-Anselmi+18} to obtain a smoothed binned covariance matrix:
\begin{equation}
    \textrm{Cov}_{X,ij} = \frac{2}{N_\sub{real} \ L^3} \int_0^\infty \frac{\de k \ k^2}{2\pi^2} \ \bar{j_0}(kr_i) \ \bar{j_0}(kr_j) \left[P_X(k)+\frac{1}{\bar n}\right]^2\, ,
    \label{eq:covmat}
\end{equation}

where $1/{\bar n}$ is the Poisson shot-noise term, $L$ is the box size of the simulations, $N_\sub{real}$ is the number of realizations (since our observed points are the mean correlation functions) and $\bar{j_0}(x)$ is a band-averaged spherical Bessel function.
In particular, if a bin is centered on $r$ and its edges are $(r_1, r_2)$:
\begin{equation}
    \bar{j_0}(kr) = \frac{\left . r^2 j_1(kr)\right|^{r_2}_{r_1}}{r^2k \Delta r \left[1+ \frac{1}{12}\left(\frac{\Delta r}{r}\right)^2\right]}
    \label{eq:Bessel_bin}
\end{equation}

with $\Delta r = r_2-r_1$, and $j_1(x)$ is the 1st-order spherical Bessel function.

The shot-noise term is equal to $L^3/N_\sub{c}$ in the case of CDM, while for halos it is taken to be $L^3/\bar N_\sub{h}$, where $\bar N_\sub{h}$ is the average number of halos at the single snapshot considered (see Table \ref{tab:halos_in_sims}).
On the other hand, Eq.~(\ref{eq:covmat}) contains the power spectrum of the tracer $P_X(k)$, which is \textit{a priori} unknown.
To avoid recomputing the covariance matrix at each step, we adopt the following prescription.
The only parameter that plays an important role in the covariance is the bias factor.
Therefore, when $X={\rm c}$ we use the linear CDM power spectrum, while for $X={\rm h}$ we assume a simple linear-bias model $P_\mathrm{h}(k)=b^2 P^\mathrm{lin}_\sub{c}(k)$.
The bias factor is then found by fitting the halo power spectrum up to scales of $k=0.1 \ h$ Mpc$^{-1}$. To perform the fit we assumed an analytical diagonal covariance matrix considering both cosmic variance and shot noise.

In Fig.~\ref{fig:covariance_lcdm} we show the comparison between the covariance (relative to a cubic box of side 1000 Mpc/$h$), measured from the DEMNUni (left) and Quijote (right) simulations, and our prescription as described above, Eq.~(\ref{eq:covmat}).
We show here both the CDM (top) and halos (bottom) measurements only for the vanilla-$\Lambda$CDM case, but we find similar agreement also for the massive-neutrino case.
Confirming previous results \cite{Best_way_to_measure_BAO-Sanchez+08,2009MNRAS.400..851S,Covmat-Gaussian-Grieb+15,2019MNRAS.482.1786L}, we find that the Gaussian-density-field approximation reproduces remarkably well not only the diagonal terms, \emph{i.e.} the variances (red dots and diamonds for CDM and halos, respectively), but also the off-diagonal ones (blue, green and yellow points) down to scales of 40 Mpc/$h$.

Finally, the fitting setup employed to estimate the LP, which minimizes biases and systematics (for both the DEMNUni and the Quijote simulations), is selected by following the procedure developed in Refs.~\cite{Linear_point_III-method-Anselmi+17,Linear_point_IV-BAO-Anselmi+18}.
In Appendix \ref{app:numerical_systematics} we explain more in detail how we applied those methodologies to our case.



\section{Results}
\label{sec:results}

In this Section we report and discuss in detail the main results of this paper, obtained following the procedure presented in the previous Section. We first assess the impact of neutrino masses on the redshift-dependence position of the LP in linear theory. Next we focus on the LP redshift evolution under the effects of non-linear gravitational evolution. To this end, we measure the LP position from simulations and compare it against an approximate cosmology-dependent analytic model. In the following subsection, we discuss the implications of our findings when employing the LP as standard ruler for massive-neutrino cosmologies. 
Finally, we examine the impact of massive neutrinos on the LP position, quantifying the shift in the LP with respect to the vanilla-$\Lambda$CDM case.  We study whether that shift can potentially be used to constrain $M_\nu$.
We also discuss the scaling of the uncertainty of the LP with the survey volume and redshift, using the set-up employed in Ref. \cite{Linear_point_III-method-Anselmi+17} to estimate the LP positions and its error.
In Appendix \ref{app:tables}, we present a table showing the estimated LP values from our $N$-body simulations.


\begin{figure}[t]
    \centering
    \includegraphics[width=1.\textwidth]{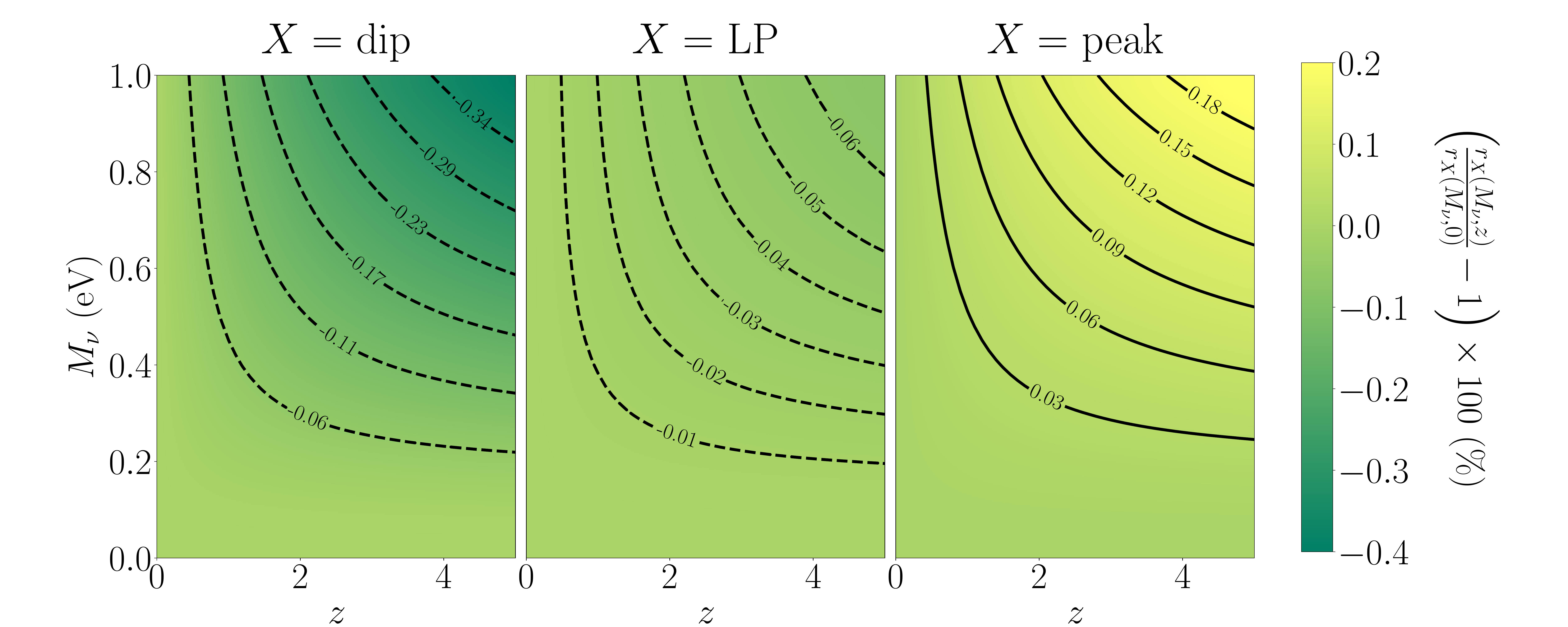}
    \caption{The evolution of the dip (left), LP (center) and peak (right) positions of the cold dark matter plus baryons TPCF in the  $z-M_\nu$ plane, according to linear theory. For each neutrino mass, the percentage difference between the quantity considered and its value at $z=0$ is plotted. Solid contour lines denote positive differences, whereas dashed lines denote negative values. Here we keep $\sigma_8$ fixed for different neutrino masses, but the result for fixed $A_\mathrm s$ is indistinguishable.}
    \label{fig:LP_linear_evolution}
\end{figure}

\subsection{Linear perturbation theory: linear point redshift evolution}
\label{sec:res:linear}

To illustrate the impact that massive neutrinos already have at linear level, in Figure \ref{fig:LP_linear_evolution} we plot the percentage difference on the position of the dip (left panel), LP (central panel) and peak (right panel) for CDM- plus-baryons TPCF compared to the same quantity computed at $z=0$ for different neutrino masses, assuming linear theory.
Different cosmologies have the same $\sigma_8$, but an almost identical result would have been obtained by fixing $A_\mathrm s$ \footnote{In Figure \ref{fig:LP_linear_evolution} we keep fixed $\sigma_8$ and $\Omega_\mathrm {m}=\Omega_\mathrm c+\Omega_\mathrm {b}+\Omega_{\nu}$. Notice however that the LP redshift-independence does not depend on which parameters we choose to keep fixed.}.
Solid and dashed contour lines represent positive and negative differences, respectively.
We stress that in absence of massive neutrinos, \emph{i.e.} with a scale-independent growth factor, the peak, dip and LP positions will be redshift-independent.} A second very important thing to notice is that the LP position is much more stable than the positions of the dip and the peak, so it is a better standard ruler. Overall, the LP shift is much smaller than the 0.5\% intrinsic uncertainty found in Ref. \cite{Linear_point_I-theory-Anselmi+15}.


\subsection{Non-linear gravity: linear point redshift evolution}
\label{sec:LP_nonlinear}

\begin{figure}[!t]
\centering
\includegraphics[width=\textwidth]{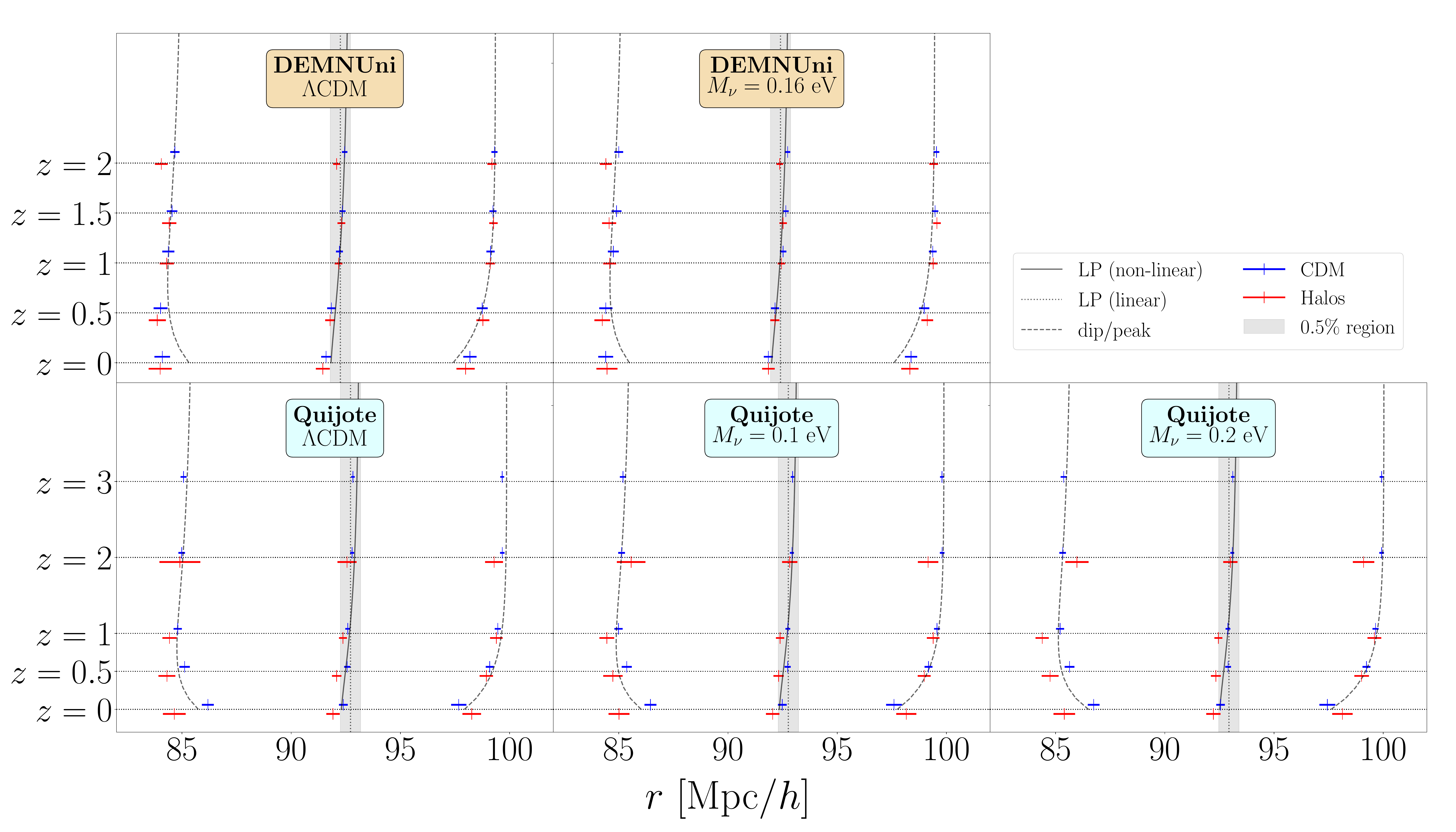}
\caption{We plot 68\% confidence limits on the position of the dip, the peak and the LP of the TPCF for every snapshot of our simulation sets.
The top panels refer to the DEMNUni set, for which we have the vanilla-$\Lambda$CDM model (top left) and the massive neutrino model (top center).
The bottom panels refer to the Quijote set with its three different models: vanilla-$\Lambda$CDM (bottom left), $M_\nu = 0.1$ eV (bottom center) and $M_\nu = 0.2$ eV (bottom right).
For each subpanel, dotted vertical lines represent the linear-theory prediction of the LP. The gray area shows the $\pm0.5\%$ LP intrinsic-bias range identified in \cite{Linear_point_I-theory-Anselmi+15} (see main text).
The remaining solid and dashed lines show the evolution in redshift of the LP, the dip and the peak, respectively, according to Eq.~(\ref{eq:TPCF_nl}). 
Blue and red bars refer to the results for CDM and halos, respectively.
Small offsets with respect to the snapshot redshifts have been introduced for the sake of clarity.
In the Quijote sector, the red bars relative to the $z=3$ snapshot are missing because the high shot noise made it impossible to have a clear estimate of the LP.
It can be noticed that the LP is particularly stable and always in agreement with the linear prediction at the 0.5\% level both for vanilla-$\Lambda$CDM and when massive neutrinos are included.
}
\label{fig:results_LP}
\end{figure}

Let us start analyzing our results from Fig.~\ref{fig:results_LP}.
For the two simulation sets (DEMNUni in the top panels; Quijote in the bottom panels) and for each model (vanilla-$\Lambda$CDM in the left panels, and massive neutrinos in the central and right panels) we plot the position of the dip on the left, the peak on the right, and the LP in the center as a function of redshift, with their 68\% relative uncertainty. As explained in Appendix \ref{app:numerical_systematics}, in order to minimize the numerical systematics, the LP estimation was performed on all the DEMNUni realizations and on the first 100 Quijote ones, fitting an 8-th degree polynomial on a range of scales spanning from 77 to 107 Mpc/$h$ for CDM and from 75 to 115 Mpc/$h$ for halos.
Blue crosses correspond to the measurement carried out on CDM field, while red ones refer to halos.
For the sake of clarity, we introduce a little offset with respect to the actual redshift of the snapshot.
In the Quijote set, red crosses at $z=3$ are missing because the low number density of halos (and the consequent high value of the shot noise) prevents us from obtaining an accurate measurement of the LP and its uncertainty.
In each subplot, the vertical dotted line represents the LP position according to linear theory.
We also compare our measurements to the values predicted by a simple non-linear model.
The non-linear TPCF can be modelled through 
Lagrangian Perturbation theory
, where the dominant effect is given by the smoothing due to the displacements from the initial positions. This approximation was already used in Refs. \cite{BAO_peak_nu-Peloso+15,noda17,2015JCAP...09..014V} and shown to reproduce well the TPCF from $N$-body simulations:

\begin{equation}
    \xi^\mathrm{nl}(r,z) = \int_0^\infty \de k \ \frac{k^2 \ P^\mathrm{lin}_\sub{c}(k,z)}{2\pi^2} \  e^{-k^2\sigma_v^2(z)} \ j_0(kr),
    \label{eq:TPCF_nl}
\end{equation}

where $P^\mathrm{lin}_\sub{c}(k)$ is the CDM linear power spectrum and

\begin{equation}
    \sigma_v^2(z) = \int \frac{\de^3 \tvec q}{(2\pi)^3} \ \frac{P^\mathrm{lin}_\sub{c}(q,z)}{q^2}
\end{equation}

is the variance of the displacement field or, equivalently, the one-dimensional velocity dispersion in linear theory.
Notice that, for our purposes, the prediction of the LP motion for CDM and halos does not change, as the substitution $P^\mathrm{lin}_\sub{c}(k) \rightarrow P_\mathrm{h}(k) = b^2 P^\mathrm{lin}_\sub{c}(k)$ only rescales the amplitude of the TPCF without shifting any scale.
Thus, the solid lines in Fig.~\ref{fig:results_LP} represent the prediction of the redshift evolution of the LP according to Eq.~(\ref{eq:TPCF_nl}), while the dashed lines do the same for the dip and the peak. The gray area shows the $\pm0.5\%$ LP intrinsic-bias range identified in Ref. \cite{Linear_point_I-theory-Anselmi+15}, i.e.~the maximum shift of the LP with respect to its linear-theory value, and the motivation for the $0.5\%$ shift in equation \eqref{eq:LP_definition}. 

Fig.~\ref{fig:results_LP} indicates that, as expected, within 1-$\sigma$, the LP position agrees at the $0.5\%$ level with the linear-theory prediction. Hence, for the vanilla-$\Lambda$CDM model, we confirm the findings of Ref. \cite{Linear_point_I-theory-Anselmi+15}, which were derived with a less rigorous analysis. More importantly, for the first time, we show that the LP position 
remains in good agreement with the linear prediction
when neutrinos are assumed to be massive. We also notice that the LP position agrees, within 1-$\sigma$, with Eq.~(\ref{eq:TPCF_nl}). Therefore, if needed,
Lagrangian Perturbation Theory
could be conveniently employed to predict the LP position.

At low redshifts, the vanilla-$\Lambda$CDM dip and peak non-linear shifts are different in the DEMNUni and Quijote sets, especially for the CDM.
Given that the cosmological-parameter values are very similar for the two sets this behavior is unlikely to be physical. A statistical fluke also seems unlikely given the agreement between the sets at high-redshift. Therefore this difference is likely due to simulation systematics: different mass resolution of the simulations, use of the approximated analytical covariance for the TPCF (\emph{i.e.} Eq.~(\ref{eq:covmat})) and numerical systematics related to the FFT-TPCF estimator not completely removed by the mitigation strategy explained in Appendix \ref{app:numerical_systematics}.
A similar trend is present also for the $\nu\Lambda$CDM case. Nevertheless, if there is a systematics difference between the DEMNUni and Quijote results, it seems to largely cancel out for the LP.

Interestingly, we find that the uncertainty on the LP position is smaller than just the average of the uncertainties on the peak and the dip: this reflects a significant anti-correlation between the latter two.
We notice that such anti-correlation almost completely disappears if we use only the diagonal part of the covariance matrix. Therefore the strong cross-correlation between different TPCF bins is responsible for the smaller uncertainty of the LP than of the peak and dip positions.


\subsection{The linear point as a standard ruler}
\label{sec:LP_standard_ruler}

In the previous Sections, we found the CDM and halo LP positions to be weakly (\textit{i.e.} ~0.5\%) affected by non-linear gravitational evolution. Along the lines of Ref.~\cite{Linear_point_I-theory-Anselmi+15}, it is crucial to also verify that redshift-space distortions do not spoil the weakness of the redshift-dependence of the LP. If that turns out to be the case, the LP can be employed as a BAO standard ruler for cosmologies where neutrinos are massive. In fact, as explained in Ref.~\cite{Linear_point_IV-BAO-Anselmi+18}, the standard-ruler properties of the LP imply that it can be used to perform Purely-Geometric-BAO (PG-BAO) distance measurements. 

BAO measurements are one of the main motivations for cosmologists to perform galaxy surveys. As detailed in Ref.~\cite{Linear_point_IV-BAO-Anselmi+18}, the PG-BAO approach allows one to estimate cosmic distances without assuming neither that the Universe is spatially flat nor a specific model for the late-time acceleration of the Universe. Furthermore the estimated distances are independent of the primordial-fluctuation parameters.

One of the consequences of our findings is that the set of cosmological models for which a PG-BAO approach has been demonstrated now includes massive neutrinos. Note that, while this holds true for the LP, it has not yet been proven for the correlation-function model-fitting approach to PG-BAO, as defined in Ref.~\cite{Linear_point_IV-BAO-Anselmi+18}.
Finally, since in this manuscript we have demonstrated the nearly redshift-independence of the LP, the next natural step will be to investigate its neutrino-mass dependence. This will inform us of its power to constrain the neutrino mass. Hence the investigation presented in Ref. \cite{2020PhRvD.101h3517O} should be extended to the massive-neutrino case.


\subsection{Linear point sensitivity to neutrino mass}
\label{sec:error-forecast}

In the real Universe, we do not know the true value of the cosmological parameters and neutrino masses.
It is, however, interesting to ask whether, assuming the $\Lambda$CDM cosmology, the LP could be used to detect the non-zero neutrino mass, if we know perfectly all the cosmological parameters.
One way to answer this question is to estimate the LP position from both real observed galaxy data and from the ``equivalent'' mock galaxy distribution where neutrinos are massless\footnote{From observed galaxy data we measure the LP in fiducial comoving coordinate \cite{Linear_point_II-BOSS-Anselmi+17, Linear_point_IV-BAO-Anselmi+18}. 
Therefore, before using the LP to detect the neutrino mass, it is required to use a fiducial cosmology close enough to the true cosmology.}.
We then ask if the two LP detections are different enough to provide a neutrino-mass detection.
In the following, we mimic this procedure using our $N$-body simulations.
We highlight that we do not aim at a very accurate answer, for which we should investigate the redshift-space distortions effects, carefully analyze how to fix all the cosmological parameters, populate our simulations with galaxies, dealing with the cross-correlation between different simulations and matching the number densities and volume of the selected galaxy surveys.
Since in this context we are not after a very accurate and detailed investigation, for simplicity we choose to apply the minimum bias setup found in Ref. \cite{Linear_point_III-method-Anselmi+17}: a 5-th order polynomial fit between 75 and 115 Mpc/$h$.
Notice finally that this exercise does not require the LP to be a BAO standard ruler. Nevertheless it has convenient properties to this end: it is weakly sensitive to non-linearities, it can be estimated in a model-independent way, it has a small uncertainty and it is independent of the fixed value of $A_{s}$ and $n_{s}$.

Our neutrino mass detector is defined by a signal-to-noise ratio (SNR) for every redshift and survey volume as the ratio between the difference of the LP position, in massive neutrino cosmologies and the massless case, and the sum in quadrature of the uncertainties in the two models.
Mathematically:

\begin{equation}
    \left. \frac{S}{N}\right|_{z, V} = \frac{r_\mathrm{LP}(M_\nu) - r_\mathrm{LP}(0)}{\left[\sigma_\mathrm{LP}^2(M_\nu) + \sigma_\mathrm{LP}^2(0)\right]^{1/2}}
           \label{eq:snr_LP}
\end{equation}

For both the vanilla-$\Lambda$CDM and the $\nu\Lambda$CDM cases, we use the LP position and error measured from $N$-body simulations. 

\begin{figure}[!t]
    \centering
    \includegraphics[width=\textwidth]{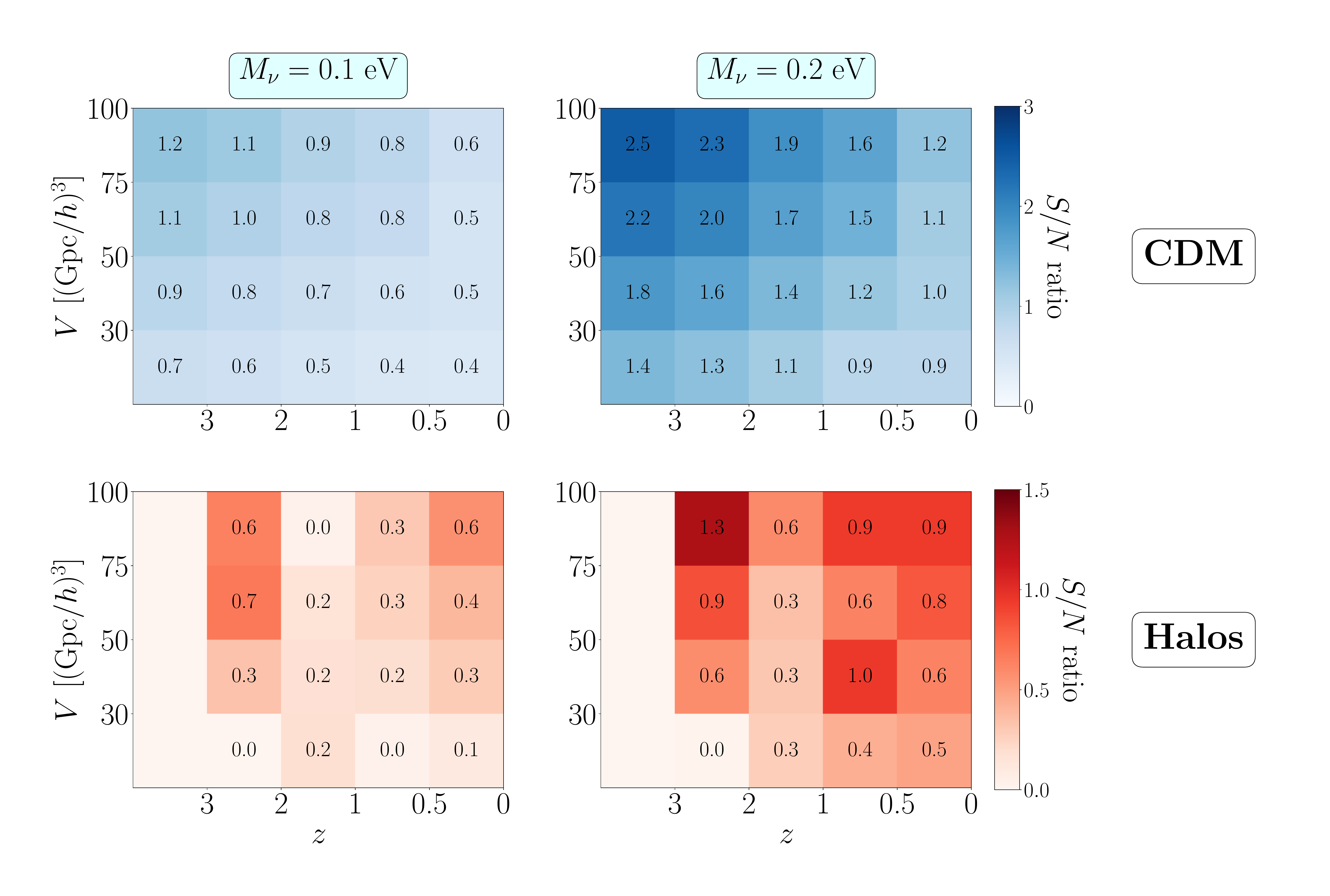}
\caption{Signal-to-noise ratio, computed with Eq.~(\ref{eq:snr_LP}), for a possible neutrino mass detection using the LP shift with respect to the vanilla-$\Lambda$CDM case. The $S/N$ caused by a neutrino mass of 0.1 (0.2) eV is displayed on the left (right) panels. Top panels show the results for CDM only, while the bottom ones show the same for halos. The left columns of the bottom panels are missing because we do not perform the analysis for halos at $z=3$.}
\label{fig:results_SNR_Quijote}
\end{figure}

The colormaps in Fig.~\ref{fig:results_SNR_Quijote} show the SNR computed with Eq.~(\ref{eq:snr_LP}) for the different models and survey volumes.
Different columns label different neutrino masses ($0.1$ eV on the left, $0.2$ eV on the right); different rows -- and different colors -- label the two different tracers,  CDM and halos, respectively.
Once again, we notice for CDM a clear trend --
the SNR increases with increasing neutrino mass, volume and redshift, for the following reasons.
First, since in the simulations $\Omega_\mathrm{m}$ is kept fixed, a larger neutrino mass means a smaller $\Omega_\sub{c}$.
This affects the shape of the CDM power spectrum, and in turn of the TPCF in a significant way at the linear level, in particular shifting the peak, the dip and the LP towards larger scales (see Fig.~\ref{fig:DEMNUni_Quijote_TPCF}).
Second, increasing the survey volume shrinks the uncertainty on the LP but does not affect its mean value: we therefore expect the denominator of Eq.~(\ref{eq:snr_LP}) to decrease for large volumes and consequently the SNR to rise.
Third, also related to the previous point, a boost of the SNR for increasing redshift is also expected.
While the LP remains stable, late-time non-linearities smear out the BAO feature;
we therefore expect the uncertainty in the zero-crossings of the first derivative of the TPCF, and thus the uncertainty in the LP, to be larger at low redshift (contextually also the second derivative, evaluated at the zero-crossings of the first derivative, should be closer to zero).
All in all, for $M_{\nu}=0.1$ eV we find the SNR to be larger than 1 when $V>30 \ \mathrm{Gpc}/h$ and $z>1$;  for $M_{\nu}=0.2$ eV the SNR is larger than 1 for most of the volumes and redshifts considered in Fig.~\ref{fig:results_SNR_Quijote}.

For halos the situation is different: contrarily to the CDM case, there is no clear trend, suggesting that our SNR is dominated by the statistical error on the LP.
Quantitatively, for $0.1$ eV neutrino mass the SNR is never greater than 1, while we only find a few cases for which this is true for  $M_\nu =  0.2$ eV.
On the other hand, the additivity of Eq.~(\ref{eq:snr_LP}) ensures that having more redshift bins can help in increasing the SNR.
For instance, from the bottom-right panel of Fig.~\ref{fig:results_SNR_Quijote}, we can see that four bins at $z=2, 1, 0.5, 0$, each of 75 (Gpc/$h)^3$ would be sufficient to detect the LP shift due to a neutrino mass of $0.2$ eV with SNR $= 2.6$. 

An important point to stress is the following.
In Section \ref{sec:LP_nonlinear} we pointed out how the uncertainty on the LP is smaller than the average of the uncertainties of the peak and the dip.
This translates to the fact that, if we reproduced Fig.~\ref{fig:results_SNR_Quijote} using the peak as our observable to detect the neutrino mass (instead of the LP), we would obtain a lower SNR because the numerical value of the denominator in Eq.~(\ref{eq:snr_LP}) will be larger than the LP case.
As a consequence using the LP position to detect the neutrino mass works better than using the peak: the LP position is found to be extremely well measured due to the high degree of anti-correlation of the peak and the dip positions.

\begin{figure}[!t]
    \centering
    \includegraphics[width=\textwidth]{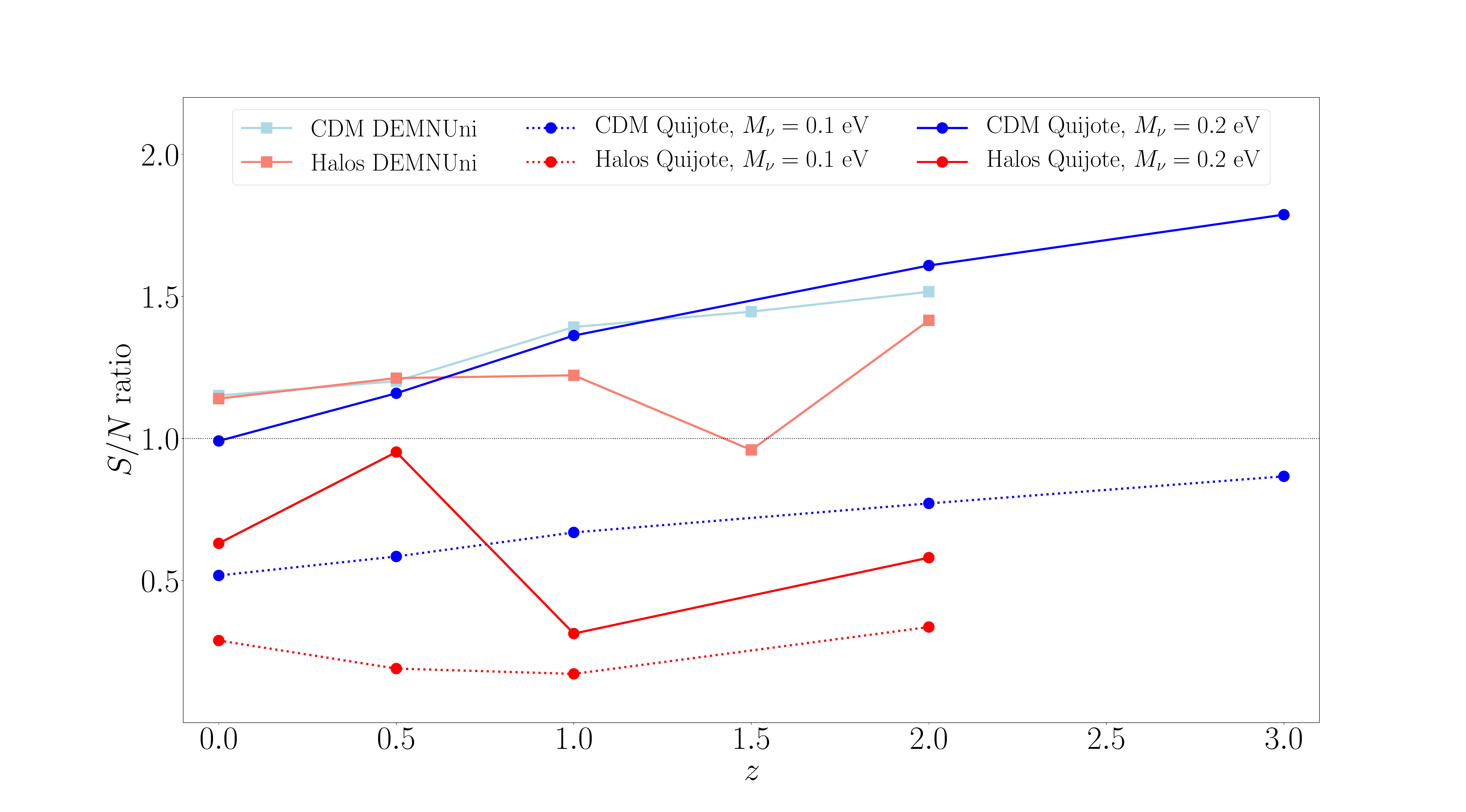}
\caption{
The figure shows the signal-to-noise ratio for a possible neutrino-mass detection using the LP shift with respect to the vanilla-$\Lambda$CDM case.
Here we compare the signal-to-noise for the 50 realizations of the DEMNUni set with the 50 of the Quijote (\emph{i.e.} our reference volume).
Squares represent the ratio for the DEMNUni set, light-blue and pink for CDM and halos respectively.
Circles refer to the Quijote set, where we distinguish the model with $M_\nu = 0.1$ eV (dashed line) and the one with $M_\nu = 0.2$ eV (dotted line).
Like in the previous figures, CDM is represented in blue, whereas halos are in red.
Finally, the dotted black horizontal line symbolizes a $S/N$ ratio equal to 1.
}
\label{fig:results_SNR_DEMNUni_vs_Quijote}
\end{figure}

In this last part, we wish to compare the SNR for the two simulations sets.
In Fig.~\ref{fig:results_SNR_DEMNUni_vs_Quijote} we show the SNR as a function of redshift.
The light-blue and pink squares are relative to the SNR as measured from the 50 DEMNUni realizations, for CDM and halos, respectively.
For the Quijote we also plot the SNR relative to 50 realizations.
As in previous plots, CDM is depicted in blue, whereas halos are in red.
Different models are represented with different line styles: dotted for $M_\nu = 0.1$ eV, solid for $M_\nu = 0.2$ eV.

Let us focus on CDM first.
Before comparing the two sets, it must be stressed that though they share the same volume, there are two main differences.
First, in the DEMNUni simulations neutrinos have a total mass of $0.16$ eV, while in the Quijote $M_\nu$ is either $0.1$ or $0.2$ eV.
Second, as already mentioned, the amplitude of fluctuations in the two sets is regulated by two different parameters: while in the DEMNUni the amplitude of the primordial scalar perturbations, $A_\mathrm s$, is constant and $\sigma_8$ decreases for increasing neutrino mass, in the Quijote $\sigma_8$ is kept fixed.
Therefore, for fixed $M_\nu$, in the latter set we expect a higher damping of the BAO feature due to non-linearities (see Fig.~\ref{fig:DEMNUni_Quijote_TPCF}) and, consequently, a higher uncertainty in the LP and a smaller SNR.
This explains why the SNR in the Quijote is much smaller than the DEMNUni for $M_\nu = 0.1$ eV but comparable with it for $M_\nu = 0.2$ eV, \emph{i.e.} for a larger neutrino mass. 

Finally, we turn our attention to halos, again underlining that there are relevant differences between the halo populations.
The different mass resolutions of the two sets imply a different minimum halo mass (see Table \ref{tab:sims}).
Table \ref{tab:halos_in_sims} shows how halos in the DEMNUni outnumber the ones in the Quijote by a factor that ranges from $\sim4-5$ at $z=0$ to $\sim 12-14$ at $z=2$.
From Fig.~\ref{fig:results_SNR_DEMNUni_vs_Quijote}, it is evident how in the DEMNUni, despite a lower neutrino mass, the SNR for a detection is substantially larger than in the Quijote, even for $M_\nu= 0.2$ eV.
This suggests that shot noise and the FFT correlation function estimator noise play a crucial role in reducing the SNR, more than neutrino mass itself.
To further support this conclusion, we notice how the evolution of the SNR in the Quijote is almost identical for different neutrino masses, as the spike at $z=0.5$ for the 0.2 eV case can still be addressed to statistical fluctuations\footnote{
We crudely estimate the standard deviation of the error on LP by assuming that the LP position is normally distributed \cite{math_stats-Kenney+51}:
\begin{equation*}
    \sigma_{\sigma_\mathrm{LP}} = \sigma_\mathrm{LP} \frac{\Gamma[(N_\mathrm{real}-1)/2]}{\Gamma[N_\mathrm{real}/2]} \sqrt{\frac{N_\mathrm{real}-1}{2} - \left(\frac{\Gamma[N_\mathrm{real}/2]}{\Gamma[(N_\mathrm{real}-1)/2]}\right)^2},
\end{equation*}
where $N_\mathrm{real}$ is the number of realizations considered.
}.
We also point out that shot noise and the bias of the tracer are tightly related to one another -- imposing a fixed-mass cut in simulations corresponds to selecting fewer halos at high redshifts, with the latter being the most massive and the most biased.

In conclusion, two different things are needed in future surveys in order to be able to detect neutrino mass using the LP: a large volume and a densely populated galaxy sample to pull down cosmic variance and shot noise. To quote some numbers reflecting the current status, we can forecast the SNR in past and upcoming surveys. One promising avenue to use the LP as a neutrino mass probe would be to rely on Intensity Mapping surveys: in this case very large volumes can be covered and the shot-noise contribution to the overall signal is expected to be small \cite{2017IMBAO}.

\begin{figure}[!t]
    \centering
    \includegraphics[width=\textwidth]{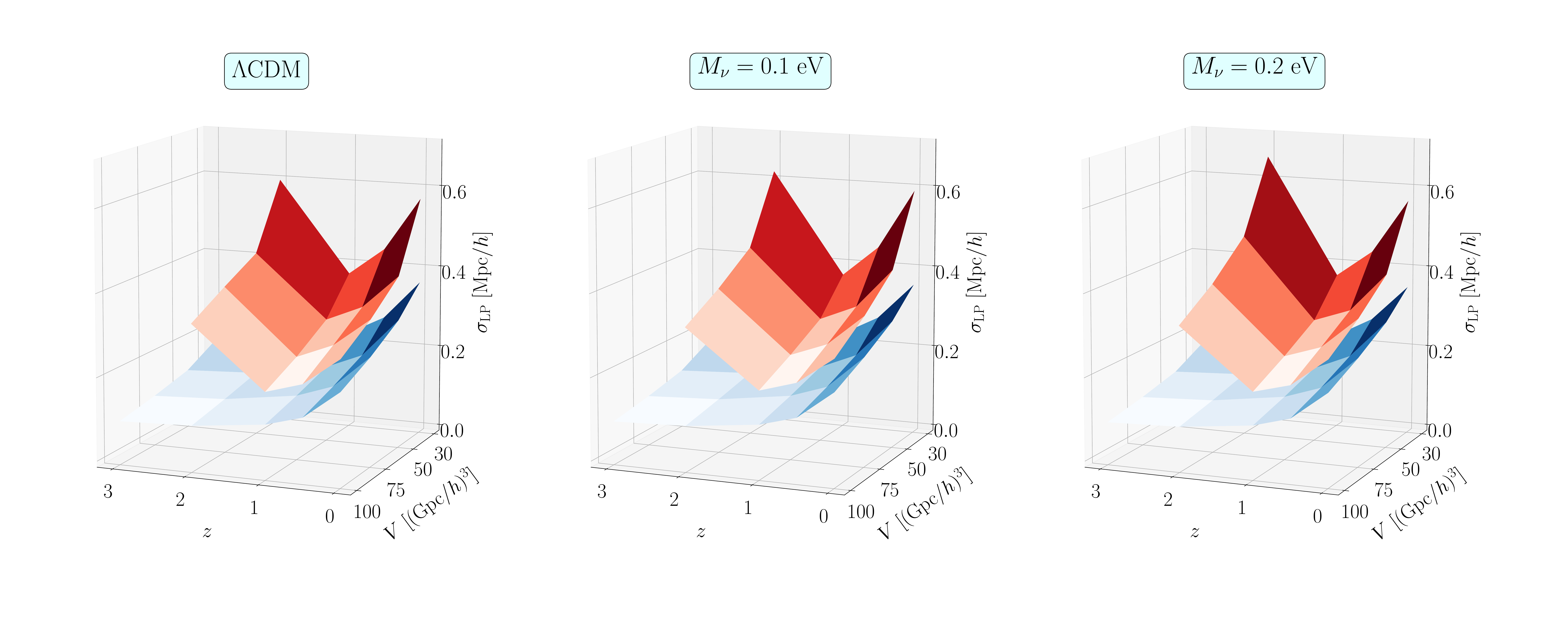}
\caption{LP position uncertainty as a function of redshift and volume (\emph{i.e.} number of realizations) in the Quijote set. We show the case for the vanilla-$\Lambda$CDM (left) and the two massive neutrino models (center and right). The blue surfaces show the 68\% uncertainty on the LP in the TPCF of the CDM, while the red ones are the equivalent for halos.}
\label{fig:results_LP_error}
\end{figure}

In order to forecast the capability of galaxy surveys to detect the neutrino mass with the LP, we need to investigate its uncertainty and how that scales with redshift, volume and number density. In Fig.~\ref{fig:results_LP_error} we plot the 68\% error on the LP position as a function of redshift and volume in the Quijote simulations.
The three different columns refer to the three different neutrino masses we consider (0 eV, $0.1$ eV, and $0.2$ eV in the left, center and right panels, respectively).
The blue surfaces show the uncertainty in the LP as measured from the TPCF of the CDM, whereas the red ones represent the same for halos.
As expected, we find a decreasing uncertainty for increasing volume, scaling as $V^{-1/2}$.
However, concerning the evolution in redshift, we notice different trends for the two tracers.
On the one hand, $\sigma_\mathrm{LP}$ for CDM monotonically increases with time due to the larger BAO smoothing caused by late-time non-linearities.
On the other hand, while we find a similar trend also for halos at late times, this tendency abruptly reverses at around $z\approx 1$.
The reason is the decreasing number density of halos at early times, and the consequent rise of shot noise, whose contribution becomes more and more important in the covariance matrix, Eq.~(\ref{eq:covmat}).

By fitting to our simulations, we obtained an empirical formula that can be used to predict the uncertainty on the LP for the tracer $X$:
\begin{equation}
    \sigma_{\mathrm{LP},X}(z,V) \approx \left[\frac{2 \ (\mathrm{Gpc}/h)^3}{V}\right]^{1/2} \left[(1+\alpha_X) \ b^{1/2} D_1^2(z) + (1+\beta_X) \ \frac{\bar n_\sub{c}}{\bar n_X(z)}\right]  \ \mathrm{Mpc}/h\,.
    \label{eq:LP_uncertainty_fit}
\end{equation}
Here we have normalized the shot-noise term with respect to the (constant) average CDM number density $\bar n_\sub c$; $b$ is the bias factor; $D_1(z)$ is the scale-independent linear growth factor; and $\alpha_X$ and $\beta_X$ are two free parameters that represent respectively deviations from the standard growth, and the shot noise $1/ {\bar n(z)}$.
The values of the parameters we obtain are quite independent of the neutrino mass, but do depend on the tracer: in particular, we find $\alpha_\sub h\approx 0.7$, $(1+\beta_\sub h)\sim 10^{-4}$ for halos, and $\alpha_\sub c \approx 0$, $(1+\beta_\sub c) \approx 0.52$ for CDM
\footnote{To find the numerical value of $\alpha_\sub h$ and $\beta_\sub h$ we use the approximation introduced in the previous footnote to estimate the uncertainty on the uncertainty on the LP.}. Notice that to forecast the galaxy-survey SNR we will use Eq.~(\ref{eq:LP_uncertainty_fit}) in parameter regions where it has not been validated. We are aware of the potential systematic error introduced by this extrapolation. However, this is adequate for our purposes here of conducting a preliminary investigation of the utility of the LP for neutrino mass detection. For the same reason, to compute the numerator of Eq.~(\ref{eq:snr_LP}) we always use the value of the LP estimated from 50 simulation boxes. In fact we found that, as expected, the LP best fit is weakly sensitive to the simulation volume considered.

Using the first and third bins of the BOSS DR-12 data \cite{BOSS-Galaxies+17} we obtain a SNR of $0.8$ for $M_\nu = 0.1$ eV and of $1.9$ for $M_\nu = 0.2$ eV.
Concerning future surveys, by assuming the 4 redshift bins, number densities, volumes and biases reported in Ref. \cite{Euclid_forecast+19}, and using the fitted values for $\alpha_\sub h$ and $\beta_\sub h$, we forecast an overall SNR $\approx 3.9$ for $M_\nu = 0.1$ eV, and SNR $\approx 5.2$ for $M_\nu = 0.2$ eV. We should finally remark that comparing the outcomes of Eq.~(\ref{eq:LP_uncertainty_fit}) to the accurate LP error estimates found for BOSS \cite{Linear_point_III-method-Anselmi+17} and Euclid \cite{Linear_point_IV-BAO-Anselmi+18} we notice that Eq.~(\ref{eq:LP_uncertainty_fit}) consistently underestimates the LP uncertainties. We warn the reader that the SNR we found for BOSS and Euclid is probably an upper limit on the results of a dedicated analysis.


We would like to stress that these results were obtained by assuming perfect knowledge of the underlying cosmology and of the fitting parameters $\alpha_\mathrm h$ and $\beta_\mathrm h$.
When allowing these parameters to vary, the SNR analysis just exposed may drastically change due to the possible degeneracies among parameters that would come into play.
Further work is the needed to provide more accurate results, also including the effect of redshift-space distortions and using galaxies instead of halos.
All in all, we expect the detection of neutrino mass from the LP shift to be particularly challenging for upcoming surveys.


\section{Conclusions}
\label{sec:conclusions}

Current BAO analyses are used to infer constraints on cosmological parameters, with particular interest in $\Omega_\sub m$ and $H_0$.
Unfortunately, this procedure might involve some problems.
The analysis is typically done by fitting the TPCF using a template function in which the cosmological parameters themselves are kept fixed and the damping factor is estimated from mock galaxy catalogs assuming a flat-$\Lambda$CDM model.
Therefore the error on the inferred distances cannot be trivially propagated to non-standard-cosmology scenarios (e.g. non-flat geometries and evolving dark energy).
Although this method has been shown to return unbiased estimates of the cosmic distances, Ref. \cite{Linear_point_IV-BAO-Anselmi+18} has showed that it may underestimate their uncertainties by up to a factor of 2.

The linear point, first proposed as a new standard ruler in Ref. \cite{Linear_point_I-theory-Anselmi+15}, can circumvent these problems.
It is insensitive to late-time non-linearities such as non-linear gravity, scale-dependent halo bias and RSD at the 0.5\% level.
Moreover, using a simple polynomial function to fit the TPCF allows a straightforward and correct propagation of uncertainties on the distances and in turn on the cosmological parameters.

In this paper we extended previous work in the vanilla-$\Lambda$CDM paradigm, to address the impact of massive neutrinos on the LP.
The goal was to check whether neutrinos affect the LP position at linear and non-linear level.
We could thus learn how to exploit the LP to infer constraints on $M_\nu$. 

We first showed, employing a Boltzmann code, that the LP position is both sensitive to the neutrino mass and redshift-independent in linear perturbation theory.
We then investigated the LP redshift evolution when we switch on the non-linear gravitational evolution. To this end, we employed two sets of state-of-art $N$-body simulations, the DEMNUni \cite{DEMNUni-Carbone+16} and the Quijote \cite{Quijote+19}.  We found the LP position to be weakly sensitive (at the 0.5\% level) to gravitational non-linearities, so its comoving position is nearly redshift-independent. 

As already done for the vanilla-$\Lambda$CDM case, we need to verify numerically that the weak redshift-dependence of the LP is not spoiled by redshift-space distortions also in massive-neutrino cosmologies.
Fortunately, our analytic understanding of why redshift-space distortions had little impact for massless neutrinos \cite{Linear_point_I-theory-Anselmi+15} carries over to the massive-neutrino case, as we theoretically checked by expanding our model of Eq. (\ref{eq:TPCF_nl}) in redshift-space.
Another effect that could pollute the standard ruler nature of the LP is the so-called growth-induced scale-dependent bias (GISDB): the small-scale suppression in the matter power spectrum caused by the presence of massive neutrinos also induces a scale- and redshift-dependence to the halo bias.
This effect was first introduced in Ref. \cite{GISDB-Parfrey+11}, subsequently studied in e.g Refs. \cite{GISDB-LoVerde1+14,GISDB-LoVerde2+14,2018Munoz}, detected in simulations in Ref. \cite{Chiang+19} and its importance highlighted in \cite{GISDB}.
The GISDB is intrinsically present in our simulation sets and from an analytical point of view, it can be modelled through a fitting function \cite{2018Munoz,Chiang+19} and we checked that using these formulae, once again, the LP motion in redshift is well within the 0.5\%.
While we leave this numerical investigation for the future, the results we found are the first important steps to show the LP can be employed as a cosmological standard ruler in this context.
Hence with the LP we can measure cosmological distances independent of the primordial cosmological parameters, of assumptions on the spatial curvature of the Universe, and of the specific model employed to describe  late-time acceleration. Our findings imply that the distances measured with the LP can be employed to constrain not only the dark energy, dark matter and baryon energy densities but also the neutrino masses. This subject deserves further investigations along the lines of Ref.~\cite{2020PhRvD.101h3517O}. 

In the last part of this work we investigated the sensitivity of the LP position to the neutrino mass value, keeping the other cosmological parameters fixed, and explored whether a detection of $M_\nu$ is possible using the shift of the LP with respect to the vanilla-$\Lambda$CDM case at a given redshift, survey volume and galaxy number density.
In this regard we did not employ the LP as a BAO standard ruler to measure cosmic distances; rather we exploited its convenient properties, in particular its weak sensitivity to non-linearities, its small uncertainty and its model-independent estimation. We proposed to compare the LP measured from real data to the one estimated from an ``equivalent'' mock galaxy distribution that assumes massless neutrinos.
Recently, Ref. \cite{BAO_phase_shift} made the first claim of a neutrino-induced phase shift of the BAO in the BOSS DR12 galaxy power spectrum.
However, in this and a related analysis \cite{2018JCAP...08..029B} several non-linear effects are incorporated using phenomenological models of the non-linear TPCF, with the inherent risk of being subject to the limitations of template-based BAO analyses (see e.g~\cite{Linear_point_IV-BAO-Anselmi+18, 2020PhRvD.101h3517O}).
In this regard the LP could provide a different route to detecting the neutrino mass.

We found that for underlying CDM field, the neutrino mass SNR detection increases for increasing volume, redshift and neutrino mass -- as expected.
On the other hand, from the halo investigation, we found that shot noise suppresses the SNR in a considerable way. Nevertheless deep tomographic redshift surveys with several redshift bins will help in increasing the SNR.
While the SNR we found for upcoming surveys are greater than unity, they are subject to important caveats: in computing them we assumed a perfect knowledge of the remaining cosmological parameters, fixed the parameters $\alpha_\mathrm h$ ans $\beta_\mathrm h$ to the values we obtained by fitting our simulations and ignored systematics in the measurement of the TPCF from simulations.
All these effects could suppress the SNR in a non-trivial and non-negligible way, therefore making it challenging to measure neutrino mass from the LP shift with respect to vanilla-$\Lambda$CDM.


\section*{Acknowledgements}
We would like to thank J. Mu{\~n}oz for useful discussion.
MV and GP acknowledge support from INFN INDARK PD51 grant.
MV also acknowledges financial contribution from the agreement ASI-INAF n.2017-14-H.0. 
The DEMNUni simulations were carried out in the framework of the "The Dark Energy and Massive-Neutrino Universe" project, using the Tier-0 IBM BG/Q Fermi machine and the Tier-0 Intel OmniPath Cluster Marconi-A1 of the Centro Interuniversitario del Nord-Est per il Calcolo Elettronico (CINECA).
We acknowledge a generous CPU and storage allocation by the Italian Super-Computing Resource Allocation (ISCRA) as well as from the HPC MoU CINECA-INAF.
FVN acknowledges funding from the WFIRST program through NNG26PJ30C and NNN12AA01C.
GDS was partially supported by a Department of Energy grant DE-SC0009946 to the particle astrophysics theory group at CWRU.

\bibliographystyle{unsrt}
\bibliography{bibtex}

\newpage
\appendix
\section{Polynomial fitting setup and TPCF numerical systematics}
\label{app:numerical_systematics}

In order to estimate the LP from the TPCF measured from simulations, one would ideally follow the methodology introduced in Ref.~\cite{Linear_point_III-method-Anselmi+17}. However that would require to have a large set of large-volume $N$-body simulations. Therefore, in order to select a consistent fitting setup to estimate the LP with a high-order polynomial, we adopt the strategy first developed in Ref.~\cite{Linear_point_IV-BAO-Anselmi+18}. We first generate 1000 TPCF realizations, approximating the TPCF distribution with a Multi-Normal, where the covariance matrix and the mean TPCF are given by Eq.s~(\ref{eq:covmat}) and (\ref{eq:TPCF_nl}) respectively. We then require that the distribution of $\chi^{2}_{\rm{min}}$ obtained from the polynomial fit to the 1000 TPCFs is consistent with the expected $\chi^{2}$ distribution, and that the mean LP from the distribution is unbiased with respect to the true LP value (the procedure is explained in greater detail in Ref.~\cite{Linear_point_III-method-Anselmi+17}). We select in this way the polynomial order and the range-of-scale over which the fit is performed. 

We stress that this procedure is self-consistent and free of numerical systematics that could affect the TPCF estimated from simulations. Notice that assuming the TPCF model given by Eq.~(\ref{eq:TPCF_nl}) has no impact on the LP subsequently estimated from simulations, it is only used as a toy model to validate the polynomial model-independent LP estimator.

By comparing with the direct-pair-count TPCF we found that the FFT correlation function estimator method (which we employ and describe in Section \ref{sec:tpcf}) introduces some spurious noise. The noise decreases if the number of particles increases and the TPCF bin size is the largest possible. Therefore we choose the maximum bin size allowed to estimate the LP -- as was shown in Ref.~\cite{Linear_point_III-method-Anselmi+17}, this corresponds to $3$ Mpc/$h$. Again, to mitigate the extra noise introduced by the few number of particles, for the halo distribution we employ a larger range of scales compared to the CDM field.

In Fig.~\ref{fig:results_LP} and Table \ref{tab:LP_table} we report the LP, dip and peak results found with an 8th order polynomial estimator fit over a range of scale of $30$ Mpc/$h$ for the CDM field (77-107 Mpc/$h$), and $40$ Mpc/$h$ for the halo distribution (75-115 Mpc/$h$); the TPCF bin width is $3$ Mpc/$h$. We used all the DEMNUni 50 realizations and the first 100 Quijote realizations in order to have enough halos to minimize the spurious extra noise introduced by the TPCF estimator. Using the full set of the Quijote would probably require an increase in the order of the polynomial and this is inhibited by the large bin size that we need to employ.

For less than half of the CDM simulation snapshots the p-value of the polynomial fit to the TPCF is below $0.005$. This is probably due to the residual spurious extra noise caused by the FFT correlation function estimator and by the approximation used to assess the TPCF covariance matrix, i.e.~Eq.~(\ref{eq:covmat}). The fits to the halo distributions do not have p-values below $0.005$. In fact the larger covariance of the halo distribution probably masks this issue. Further investigations of the numerical issues and convergence tests related to the FFT correlation function estimator are beyond the scope of this paper and will be the subject of future work.



\newpage
\section{Linear point position}
\label{app:tables}

In this Appendix we report the estimations of the linear point position from $N$-body simulations as presented in Section \ref{sec:LP_nonlinear} and reported in Fig.~\ref{fig:results_LP}. 

\begin{table}[!ht]
\begin{center}
\renewcommand{\arraystretch}{1.1}
\resizebox{1\textwidth}{!}{
 \begin{tabular}[t]{ l||c|c } 
  \multicolumn{3}{c}{\LARGE \textbf{DEMNUni} \normalsize} \\
  \hline \hline
    \multicolumn{3}{c}{\large  \normalsize} \\
  \multicolumn{3}{c}{\large $\textbf{vanilla-} \normalsize\boldsymbol \Lambda$\textbf{CDM} \normalsize} \\
  \multicolumn{3}{c}{\large Linear prediction: 92.25 Mpc/$h$ \normalsize} \\
  \hline \hline
  \hline
  \textbf{redshift}    & CDM & Halos  \\
  \hline \hline
    $\textbf{$z = 2$} $   & $92.45 \pm 0.12$ & $92.08 \pm 0.16$ \\
    $\textbf{$z = 1.5$} $ & $92.35 \pm 0.14$ & $92.30 \pm 0.18$ \\
    $\textbf{$z = 1$} $   & $92.21 \pm 0.16$ & $92.17 \pm 0.18$ \\
    $\textbf{$z = 0.5$} $ & $91.84 \pm 0.20$ & $91.78 \pm 0.23$ \\
    $\textbf{$z = 0$} $   & $91.60 \pm 0.22$ & $91.45 \pm 0.32$ \\
  \hline 
  
  \multicolumn{3}{c}{\large  \normalsize} \\
  \multicolumn{3}{c}{\large $\boldsymbol{\nu\Lambda}$\textbf{CDM} ($M_\nu = 0.16$ \text{eV})} \\
  \multicolumn{3}{c}{\large Linear prediction: 92.40 Mpc/$h$ \normalsize} \\
  \hline \hline
  \hline
  \textbf{redshift}    & CDM & Halos  \\
  \hline \hline
    $\textbf{$z = 2$} $   & $92.73 \pm 0.12$ & $92.37 \pm 0.15$ \\
    $\textbf{$z = 1.5$} $ & $92.65 \pm 0.13$ & $92.52 \pm 0.17$ \\
    $\textbf{$z = 1$} $   & $92.52 \pm 0.15$ & $92.45 \pm 0.17$ \\
    $\textbf{$z = 0.5$} $ & $92.15 \pm 0.18$ & $92.14 \pm 0.21$ \\
    $\textbf{$z = 0$} $   & $91.85 \pm 0.21$ & $91.85 \pm 0.29$ \\
  \hline

 \end{tabular}
 
  \hspace{0.2cm}
  
   \begin{tabular}[t]{ l||c|c } 
  \multicolumn{3}{c}{\LARGE \textbf{Quijote} \normalsize} \\
  \hline \hline
    \multicolumn{3}{c}{\large  \normalsize} \\
 \multicolumn{3}{c}{\large $\textbf{vanilla-} \normalsize\boldsymbol \Lambda$\textbf{CDM} \normalsize} \\
  \multicolumn{3}{c}{\large Linear prediction: 92.71 Mpc/$h$ \normalsize} \\
  \hline \hline
  \hline
  \textbf{redshift}    & CDM & Halos  \\
  \hline \hline
    $\textbf{$z = 3$} $   & $92.83 \pm 0.08$ & \xmark  \\
    $\textbf{$z = 2$} $   & $92.79 \pm 0.09$ & $92.56 \pm 0.44$ \\
    $\textbf{$z = 1$} $   & $92.59 \pm 0.11$ & $92.37 \pm 0.18$ \\
    $\textbf{$z = 0.5$} $ & $92.57 \pm 0.15$ & $92.08 \pm 0.22 $ \\
    $\textbf{$z = 0$} $   & $92.39 \pm 0.20$ & $91.91 \pm 0.31$ \\
  \hline 
  
    \multicolumn{3}{c}{\large  \normalsize} \\
  \multicolumn{3}{c}{\large $\boldsymbol{\nu\Lambda}$\textbf{CDM} ($M_\nu = 0.1$ \text{eV})} \\
    \multicolumn{3}{c}{\large Linear prediction: 92.77 Mpc/$h$ \normalsize} \\
  \hline \hline
  \hline
  \textbf{redshift}    & CDM & Halos  \\
  \hline \hline
    $\textbf{$z = 3$} $   & $ 92.96 \pm 0.08$ &  \xmark \\
    $\textbf{$z = 2$} $   & $ 92.93 \pm 0.09$ & $ 92.83 \pm 0.35 $ \\
    $\textbf{$z = 1$} $   & $ 92.74 \pm 0.11$ & $ 92.38 \pm 0.19 $ \\
    $\textbf{$z = 0.5$} $ & $ 92.73 \pm 0.14$ & $ 92.32 \pm 0.24$ \\
    $\textbf{$z = 0$} $   & $ 92.49 \pm 0.20$ & $ 92.05 \pm 0.31$ \\
  \hline 
  
      \multicolumn{3}{c}{\large  \normalsize} \\
  \multicolumn{3}{c}{\large $\boldsymbol{\nu\Lambda}$\textbf{CDM} ($M_\nu = 0.2$ \text{eV} )} \\
  \multicolumn{3}{c}{\large Linear prediction: 92.93 Mpc/$h$ \normalsize} \\
  \hline \hline
  \hline
  \textbf{redshift}    & CDM & Halos  \\
  \hline \hline
    $\textbf{$z = 3$} $   & $93.11 \pm 0.08$ &  \xmark \\
    $\textbf{$z = 2$} $   & $93.09 \pm 0.09$ & $ 93.00 \pm 0.33$ \\
    $\textbf{$z = 1$} $   & $92.89 \pm 0.11$ & $ 92.45 \pm 0.18$ \\
    $\textbf{$z = 0.5$} $ & $92.90 \pm 0.14$ & $ 92.34 \pm 0.23$ \\
    $\textbf{$z = 0$} $   & $92.55 \pm 0.20$ & $ 92.22 \pm 0.32$ \\
  \hline 
   \end{tabular}
 
}
  \end{center}
\caption{The table summarizes the results for the LP (including the 0.5\% correction) for the DEMNUni (left) and the Quijote (right) simulations. We show the LP position with 1-$\sigma$ uncertainty.}
\label{tab:LP_table}
\end{table}

\end{document}